\documentclass[12pt,preprint]{article} 
\usepackage{graphicx}

 \usepackage{color}           
 \definecolor{DarkGreen}{rgb}{0.0,0.45,0.0}  

\begin{document}
\title{Research progress based on observations of the New Vacuum Solar Telescope}
\author{Xiaoli Yan$^{1,2}$, Zhong Liu$^{1,2}$, Jun Zhang$^{3}$, Zhi Xu$^{1,2}$ \\ }

\date{}
\maketitle

\noindent {The purpose of this paper is to introduce the main scientific results made by the one-meter New Vacuum Solar Telescope (NVST),  which was put into commission on 2010. NVST is one of the large aperture solar telescopes in the world, located on the shore of Fuxian lake of Yunnan province in China, aiming at serving solar physicists by providing them with high resolution photospheric and chromospheric observational data. Based on the data from NVST and complementary observations from space (e.g., Hinode, SDO and IRIS, etc), dozens of scientific papers have been published with a wide range of topics concentrating mainly on dynamics and activities of fine-scale magnetic structures and their roles in the eruptions of active-region filaments and flares. The achievements include dynamic characteristics of photospheric bright points, umbral dots, penumbral waves, and sunspot/light bridge oscillation, observational evidence of small-scale magnetic reconnection, and fine-scale dynamic structure of prominences. All these new results will shed light on the better understanding of solar eruptive activities.  Data release, observation proposals, and future research subjects are introduced and discussed. 

\noindent\rule[0.1cm]{13.5cm}{0.01cm}\\
$^{1}$Yunnan Observatories, Chinese Academy of Sciences, Kunming, Yunnan 650216, People's Republic of China.(email: yanxl@ynao.ac.cn)\\
$^{2}$ Center for Astronomical Mega-Science, Chinese Academy of Sciences, 20A Datun Road, Chaoyang District, Beijing, 100012, People's Republic of Chin.\\
$^{3}$ Key Laboratory of Solar Activity, National Astronomical Observatories, Chinese Academy of Sciences, 100012, Beijing,People's Republic of China.\\
\section{Introduction of observational Systems}
NVST is a vacuum solar telescope with a 985 mm clear aperture located at the Fuxian Solar Observatory of the Yunnan Observatories[1]. The main instruments include the high-resolution multi-channel imaging system[2] and the high dispersion multi-wavelength spectrograph[3][4]. The two systems have identical focus and are arranged to be perpendicular to each other. The multi-wavelength spectrograph can observe two chromospheric lines H$\alpha$ 656.3 nm and Ca II 854.2 nm, and one photospheric line Fe I 5324, simultaneously. In this paper, we only focus on the imaging system. The working wavelengths and the main properties of each channel are summarized below.

\subsection{High Resolution Multi-channel Imaging System}
The imaging system comprises five channels, i.e., two broad-band interference filter channels and three narrow-band Lyot filter channels. {\bf The working wavelengths} and the main properties of each channel are summarized as follows:

\begin{itemize}

\item[*] H$\alpha$ channel, to monitor magnetic structures in the chromosphere. The central wavelength can be tunable in the range of 656.28 $\pm$ 0.4 nm  and the full bandpass width is 0.025 nm (e.g., giving a FWHM of an order of 11 km/s).

\item[*] Ca II H/K channel, to monitor magnetic structures in the lower chromosphere. H and K lines are alternative. The central wavelength can be tunable in the range of 393.3 (396.8) $\pm$ 0.5 nm and the full bandpass width is 0.05 nm.

\item[*] He I 1083 nm, a near infrared proxy of coronal holes, to explore magnetic fields in the low corona. The central wavelength can be tunable in the range of 1083 $\pm$ 0.5 nm and the full bandpass width is 0.05 nm.

\item[*] G band, to indicate small magnetic structures in the deep photosphere. The central wavelength is 430 nm and the full bandpass width is 1 nm.

\item[*] TiO band, sensitive to the temperature, to allow easy detection of BPs in granular lanes or umbral dots. It is centered at 705.8 nm and the full bandpass width is 1 nm.

\end{itemize}

Although only three channels including H$\alpha$, TiO and G-band have been used, the observations provide impressive evidence of the high-resolution capability of the NVST. Particularly the off-band observation in H$\alpha$ is very impressive[5]. The post-processing with the speckle masking method produces the results with the resolution close to the diffraction limit of the TiO band. There are 3 levels of data products. Level-0 represents the raw status. Level-1 is achieved based on a luck image selection algorithm, followed by the dark current and flat field modification. Level-1.5 is reconstructed by a speckle masking method[5][6][7]. To do so, we have to take a large number of short-exposure images (H$\alpha$ is 20 ms, TiO is about 1ms). Using a large number of frames (more than 100 frames) and a statistical algorithm, we are able to reconstruct one frame with the whole field of view ($\approx$ 3 arcmin). H$\alpha$ images have a pixel size of 0.$^\prime$$^\prime$163 and a cadence of 12 s. In addition, TiO images have a pixel size of 0.$^\prime$$^\prime$04. The time cadence of Level 1 and 1.5 data products is about 12 seconds in H$\alpha$ channel and 30 seconds in TiO channel.

The left panel of Fig. 1 shows an example of a complex active region (AR) named NOAA 12673 observed on 2017 September 5. The penumbra threads and fine structures in a light bridge are well resolved in the high resolution TiO image. Fine structures of the chromospheric fibrils, umbra dots, and penumbra wave can be clearly found in the observation of a sunspot at the H$\alpha$ center images. Zhu et al. (2016)[8] found that the magnetic field lines from non-linear force-free field (NLFFF) extrapolations are consistent with the patterns of H$\alpha$ fibrils observed by the NVST. In particular, the fine structure of solar filaments observed by the NVST is very complicate. The right panel of Fig. 1 shows an example of a quiescent filament observed at 05:29 UT on July 23, 2018. Many thin threads perpendicular to the spine are found. These data can be used to study the structure and evolution of different atmospheres from the photosphere to the chromosphere. 

\subsection{High Dispersion Multi-wavelength Spectrographs}

The NVST actually consists of two vertical spectrographs, which are alternative and co-existing in space by dispersing the light at two different directions. They share the same entrance slit, but the slit direction has to be changed by rotation. {\bf The length along} the slit is about 136 arcseconds. Several slits with different widths (60 - 150 micrometer) can be selected. One spectrograph works at the optical wavelength band. The focus length of the collimator is about 6 meters. A 1200 mm$^{-1}$ grating is used to obtain two chromospheric lines H$\alpha$ and Ca II 854.2 nm and one photospheric line Fe I 5324, simultaneously. In addition, the grating can be rotated to switch the observation window to other lines. The main properties of this spectrograph are listed in Table 1. The other spectrograph is for the near-infrared band. The focus length is about 9 meters. A 316  mm$^{-1}$ echelle grating is used to obtain the high order spectrum of the He I 10830 nm (the 5th order) and Fe I 1.56 micrometer (the 3rd order). The precise reduction of these spectra observed by the NVST can be seen from Wang et al. (2013), and Cai et al. (2017, 2018)[3][4][9].  
  \begin{table}
\begin{center}
\caption[]{Main properties of the Multi-Wavelength Optical-band Spectrograph. }\label{Tab:1}

 \begin{tabular}{lr}
  \hline
  \hline\noalign{\smallskip}
Spectrometer    &     blazed grating (rotatable)\\
Grating size    &     156mm$\times$220mm\\
Grooves per mm    &    1200 g/mm\\
Blazed angle [$^{\circ}$]    &    36.8  (1st order, blazed wavelength 10000 {\AA})\\
Present slit width [microns]    &   100 (0.45 arcsec) \\
Linear dispersion [mm/{\AA}]        &   0.75 @ H$\alpha$,  0.82 @ Ca II 8542,  2 @ Fe I 5324\\
Raw spectra frame size [pixels]   &   2672$\times$4008\\
Pixel size [microns]            &         9$\times$9\\
Spectral sample [m{\AA}/pixel]    &   12 @ H$\alpha$,  11 @ Ca II 8542, 4.5 @ Fe I 5324\\
Spatial sample [arcsec /pixel]    &    0.041\\
\noalign{\smallskip}\hline
\end{tabular}
\end{center}
\end{table}

\section{Data release and Proposed observations}

The NVST has been taking the daily observations since Oct. 2012. Quick-look movies (Level1 or Level1+) of all observation entries can be browsed at http://fso.ynao.ac.cn. Data requirement can be submitted through this website. In addition to the daily observations, the NVST also takes observations proposed by scientists around the world. Applications for observation proposals usually start around the middle of May and end in the middle of June. Application form can be downloaded from http://fso.ynao.ac.cn/proposal.aspx. The contents related to terminals may change every year according to the operation condition. Scientists affiliated to the CAS can access the application form by China Virtual Observatory (China-VO). The NVST scientific committees will review and evaluate all proposals. The accepted proposals will be carried out at the NVST from September to November. The principle investigator (PI) has the priority to keep the observation data for three months. After three months, the data will be open on our website.

Since 2015, the NVST has carried out many proposed observations from National Astronomical Observatories of China, Purple Mountain Observatories, Peking University, Shandong University, Xinjiang Observatories, National Space Science Center, Kunming University of Science and Technology, Hida observatory of Japan, and Indian Institute of Astrophysics. 
In addition, the NVST has also taken co-ordinate observations with other space or ground-based telescopes, such as GST at BBSO, Domeless Telescope at Hida, Hinode, IRIS, and GREGOR.

\section{Main scientific results based on the NVST data}
 High Resolution Multi-channel Imaging System can image the photosphere and the chromosphere simultaneously. Based on these high resolution data, many features in the photosphere and the chromosphere were studied. These results are divided into five parts according to their different properties.

\subsection{Dynamic characteristics of photospheric bright points, umbral dots, penumbral wave, and sunspot/light bridge oscillation}
{\bf Since the NVST started observations, many quiet regions and sunspots have been observed.} In quiet sun, many dynamical features can be seen, such as the bright points, granules, granule dark lane, and so on. The light bridges and umbral dots are often seen in the umbra of sunspots. In the upper atmosphere of sunspots, especially in the chromosphere, penumbral wave and umbral oscillations are the main features. Using the high resolution data of the NVST, the following studies were carried out.

\subsubsection{Bright points}
In high-resolution observation of photosphere, bright points are prominent features located between the granules. Using six high-resolution TiO-band image sequences taken by the NVST from 2012 to 2014, Ji et al. (2016a)[10] investigated the morphologic, photometric dynamic properties of BPs in terms of equivalent diameter, the intensity contrast, lifetime, horizontal velocity, diffusion index, motion range, and motion type. In this study, the Laplacian and morphological dilation algorithm (LMD)[11] and three-dimensional segmentation algorithms were used to detect and trace the evolution of  the BPs. Moreover, the statistical properties of the magnetic fields of BPs are also explored and compared by using the vector magnetograms.

Many useful parameters of BPs are obtained as follows: The areas of BPs range from 0.2\% to 2\%; The mean equivalent diameters range from 168$\pm$29 $\rm km$ to 195$\pm$36\,$\rm km$; The mean ratios of intensity contrast range from 0.99$\pm$0.04 to 1.06$\pm$0.05; The mean lifetimes range from 104 \,$\rm s$ to 141\,$\rm s$; The mean horizontal velocities range from 1.04$\pm$0.54 km $s^{-1}$ to 1.35$\pm$0.71\, km $s^{-1}$; The mean diffusion indices range from 0.86$\pm$0.39 to 1.31$\pm$0.54; The mean ratio of motion range values range from 0.96$\pm$0.67 to 1.30$\pm$0.80, and the mean index of motion type values range from 0.58 to 0.69. Note that the rate of motion range is defined as $m_r=\sqrt{{X_{max}-X_{min}}^2+{Y_{max}-Y_{min}}^2}/r$, where $X_{max}$ and $X_{min}$ are the maximum and minimum coordinates of the path of a single BP in the x-axis, and $Y_{max}$ and $Y_{min}$ are the ones in the y-axis, r is the radius of the circle which corresponds to the maximum size of the BP during its lifetime. The index of motion type is defined as $m_t=d/L$, where $d=\sqrt{{X_n-X_1}^2+{Y_n-Y_1}^2}$ and L is the whole path length. In addition, the relationship between the properties of BPs and their embedded magnetic environments has been investigated. It is found that the stronger magnetic field corresponds to the bigger and brighter BPs, while the weaker magnetic field corresponds to the smaller and weaker, faster diffusion, and faster movement of the BPs. The detailed parameters are shown in table 2.

Liu et al (2018)[12] studied the properties of isolated BPs[13] and non-isolated BPs in an active region observed at TiO (705.8nm) wavelength by the NVST with the seeing parameter r0 of 10.628 $\pm$ 0.220 cm. They presented a novel algorithm of identifying and tracking BPs by using the Laplace kernel and the second order derivatives in x- and y-directions to obtain the edge information, and by assuming that two BPs obtained in single frames in consecutive images is the same one appearing at different time if one has overlapped with the other. Based on these ideas, they identified and tracked the photospheric BPs, and further classified them into the isolated (see Figs. 2(a)-2(e)) and non-isolated ones (Figs. 2(f)-2(j)). The isolated BPs are the ones that did not experience split and mergence in their lifetimes. However, the non-isolated ones are those that experienced split and mergence for at least once in their lifetimes. In the algorithm, an evolving BP is considered as a non-isolated BP if it appeared in one frame as a single blob in a frame but more than one blob appeared in previous or next consecutive images. Otherwise, the evolving BP is considered as an isolated one.

Once the BPs were identified, tracked, and classified, they started to study the isolated and the non-isolated BPs in three regions with different background magnetic fields. Namely, Region 1 contains a small pore with the strongest background fields; Region 2 has no pores but contains elongated BPs with weaker background fields; Region 3 contains the least BPs with the weakest background fields. In these regions, they found that the number density of BPs obtained in single frame appeared most in Region 1, and least in Region 3. Therefore, the number density of snapshot BPs was considered to be strongly dependent on the strength of the background fields as well as the area coverage of BPs. On the other hand, the size/intensity contrast distributions of both the isolated and the non-isolated BPs were found to be independent of the strength of the background fields. However, their lifetimes are slightly shorter in the stronger background filed regions. 

BPs are located in the dark inter-granular between the granule cells. By using the data of GST, Ji et al. (2012)[14] found that there are many ultrafine channels rooted in the dark inter-granular connecting from the photosphere to the corona, in which the energy may be transferred from the lower atmosphere to the higher one. Whether BPs play an important role in the energy transfer or not is deserved for further investigation.

\subsubsection{Umbral dot, penumbral wave}
Umbral dots (UDs) are small isolated brightenings in sunspot umbrae under high resolution observations. Their appearance indicates the existence of convection. According to their locations inside an umbra, UDs can be divided into two types: central UDs (CUDs) and peripheral UDs (PUDs). To investigate UD properties, the difference between UDs and PUDs, and the relationship between their dynamic properties and magnetic field strengths, Ji et al. (2016)[15] selected high-resolution TiO images of four ARs taken by the NVST. Six sunspots were selected from the four ARs (NOAA 11598, 11801, 12158, and 12178). A total of 1220 CUDs and 603 PUDs were identified by using the phase congruency technique [16]. Meanwhile, the line-of-sight magnetograms of the sunspots taken with the Helioseismic and Magnetic Imager (HMI) on-board the Solar Dynamics Observatory (SDO) were used to determine umbral magnetic field strengths. The main results are as follows: The diameters and lifetimes of UDs increased with the brightness, but velocities did not. Moreover, diameters, intensities, lifetimes and velocities depend on the surrounding magnetic fields. The CUD diameter was larger, brighter, and has longer lifetime and slower motion in a weak umbral magnetic field environment than that in a strong one. By using Goode Solar Telescope (GST) data, Su et al. (2016)[17] also found that the wavefronts rotate the counter-clockwise and clockwise rotation alternately in the sunspot umbra.

Typical 5-minute and 3-minute oscillations were found in the photosphere and chromosphere of the Sun for dozens of years. High resolution GST observation also revealed that the periods of the running umbral waves are from 2.2 to 2.6 minutes [17][18]. Using high spatial and temporal resolution observations of the NVST and Atmospheric Imaging Assembly (AIA) on board the SDO, Wang et al. (2018)[19] discovered one minute oscillation at different heights above sunspot umbra. They used a novel time-frequency analysis method named the synchrosqueezing transform (SST) to represent their power spectra and to reconstruct the high-frequency signals at different solar atmospheric layers. The method SST is capable of resolving weak signals even when their strength is comparable to the high-frequency noise. A significant enhancement between 10 mHz and 14 mHz (labeled as 12 mHz) at different atmospheric layers was found by analyzing the spectrum in the umbra. Moreover, the 12 mHz component exists only inside the umbra.  Besides, Zhou et al. (2017a, 2017b)[20][21] found that the period of the running waves within the umbra is half of that within the penumbra. The former is a typical 3-min oscillation with a period of 156 s and the latter  is a typical 5-min oscillation with a period of 312 s.

\subsubsection{Light wall on the sunspot light bridge}
Using high resolution photospheric and chromospheric data from the NVST, Yang et al. (2015)[22] found a special structure rooted in the light bridge of the sunspot and this structure exhibited upward and downward movements. They named this structure as light-wall. The deprojected mean height, amplitude, oscillation velocity, and the dominant period are determined to be 3.6 Mm, 0.9 Mm, 15.4 km $s^{-1}$, and 3.9 minutes, respectively. They interpreted the oscillations of the light wall as the leakage of p-modes from below the photosphere. In the following, Hou et al. (2016)[23] reported that the light wall can also appear near the umbral-penumbral boundary and along a neutral line between two small sunspots. Moreover, the wall body consists of multi-layer and multi-thermal structures that occur along magnetic neutral lines in active regions. In the following, Tian et al. (2018)[24] found that the surge-like activity previously reported above light bridges in the chromosphere has two components. One component is the ever-present short surges likely to be related to the upward leakage of magneto-acoustic waves from the photosphere. The other is the occasionally occurring long and fast surges that are obviously caused by the intermittent reconnection jets.

\subsection{Formation, fine-scale structures, and eruption mechanism of solar filaments}
Solar filaments/prominences are very common features in the solar atmosphere and relatively dense and cool plasma structures embedded in the low-density and hot corona. Due to the different locations of their appearances, they are given different names. When seen in emission above the limb, they are called as prominences, and they are called filaments when seen in absorption line ($H\alpha$) against the solar disk. According to their locations on the solar disk, they can be divided into three types, i.e., quiescent filaments, active-region filaments, and intermediate filaments. Quiescent filaments are long-lasting structures with lifetimes of several days, weeks, or even much longer. Compared with quiescent filaments, active-region filaments are relatively short-lived structures lasting for several hours or days. The eruption of solar filaments is often associated with solar flares and coronal mass ejections (CMEs). A typical coronal mass ejection (CME) includes three parts, i.e., bright front, dark cavity, and bright core. It is generally believed that the bright core of a CME is a filament [25]. Therefore, studying solar filaments is very important for understanding the productions of CMEs. 

\subsubsection{Fine structure of solar prominences}
The complex structures of solar prominences have always attracted people's attention. In general, quiescent prominences have two linked categories of structures: ``spine" and ``barb". A``spine" is the longest part of a quiescent prominence that forms a horizontal axis composed of many resolvable threads, while``barb" is connected to the spine and terminates in the chromosphere[26]. 

Due to the good seeing in Fuxian lake, a lot of quiescent prominences were observed by the NVST. The ubiquitous counter-streaming flows are the most prominent features in the prominences. Yan et al. (2015)[27] found that the large-scale counter-streaming flows with a certain width in a solar prominence are co-aligned well with the prominence threads on 2012 November 2 (see Fig. 3(a)). The velocity of these material flows ranged from 5.6 km $s^{-1}$ to 15.0 km $s^{-1}$ (see Figs. 3(b)-(d)). In another quiescent prominence, the formation of several parallel tube-shaped structures in the barb is found and the width of the structures ranged from about 2.3 Mm to 3.3 Mm during the evolution of the prominence on 2013 September 29. The parallel tube-shaped structures merged together accompanying with material flows from the spine to the barb. Li et al. (2018)[28] found that the material in the middle section of a prominence is mainly restricted to flow back and forth at a certain region. Shen et al. (2015)[29]  have analyzed the evolution of a bubble structure and the counter-streaming flows in a prominence on 2014 May 20 in detail (see Fig. 4a). They found that the collapse and oscillation of the bubble boundary were tightly associated with a flare-like feature located at the bottom of the bubble. To obtain the velocities of the upward and downward flow in the prominence, the time-distance diagrams were made along the curve lines A1, A2, B1, and B2 in Fig. 3a. The average speeds of the upward and downward flows along the curve lines A1 and A2 are 13.7 $\pm$ 5.2 km $s^{-1}$ and 12.5  $\pm$ 5.2 km $s^{-1}$, respectively (see Figs. 4b and 4c). The average speeds of the counter-streaming flows in the spine of the prominence along the curve lines B1 and B2 are 20.1 $\pm$ 4.5 km $s^{-1}$ and 15.8 $\pm$ 3.8km $s^{-1}$, respectively (see Figs. 4d and 4e). 

Except for the counter-streaming flows, the Kelvin-Helmholtz instability was also identified in a quiescent prominence by Yang et al. (2018) and Li et al. (2018)[30][31]. Using the NVST red-wing H$\alpha$ data on 2016 June 4 and 2017 September 18, the vortex-like structures were detected on the interface/surface of prominence plumes by Yang et al. (2018) and Li et al. (2018). Yang et al. (2018) found 2$^\prime$$^\prime$ vortex before the formation of 5$^\prime$$^\prime$ size vortex in the prominence. Moreover, the 255s oscillatory period was identified in the prominence. Li et al. (2018) measured the speeds of the mass flows in the prominence and found that they are larger than the local sound speed.  They suggested that the kelvin-Helmholtz instability may occur in the interface/surface of prominence plumes. The vortex-like structure was also found during the evolution of the prominence (see H$\alpha$ off-band observation in Fig. 5(a)-(c)). The line-of-sight velocity of the vortex-like structure exhibited red shift (see Fig. 5(d)). The thread oscillation is another characteristic of prominences. Due to the high spatial and temporal resolution data of the NVST, two types of the small-scale oscillations of prominence threads are found by Li et al. (2018). In one type, the oscillation period firstly decreased and then increased. In the other type, the oscillation period grew quickly at the beginning and then decreased. Note that the direction of these small-scale oscillations mentioned above is perpendicular to that of the prominence threads.

The nature of the complex magnetic structures and the mass flows in the prominences is still unclear. The formation mechanisms and driven forces of the mass flows are deserved for further investigation. In addition, it is very difficult to explain the complex magnetic structures of solar prominences observed by the NVST.  Therefore, there is still a long way to completely understand the magnetic structures and mass flow of solar prominences.

\subsubsection{Filament formation}
Due to the low time resolution data in the past, the formation process of solar filaments is difficult to capture. Up to now, formation and eruption mechanisms of solar filaments are still controversial. Especially, the studies on the formation of filaments are very rare. Thanks to the high temporal and spatial resolution observation of the NVST, several cases of filament formation were captured by the NVST. Based on these observations, several new formation mechanisms of solar filaments were proposed.

Yan et al. (2015)[32] studied the formation and magnetic structures of two successive active-region filaments in AR NOAA 11884. From 2013 October 31 to November 3, there are four-day observations with nice seeing. During this period, there are two active-region filaments formed successively at the same position. Before the filament formation, there are several chromospheric fibrils connected a small sunspot with negative polarity (see Fig. 6a). The green arrow points to the small sunspot in Figs. 6a and 6b. Through studying the vector magnetograms and TiO observations, the small sunspot was found to rotate around its center in the photosphere. Due to the small sunspot rotation and the shearing motion between the two opposite polarities, the first filament formed at the polarity inversion line (PIL) (see Fig. 6b). The yellow arrow in Fig. 6b indicates the newly formed filament. Using the non-linear force free field extrapolation (NLFFF) based on the vector magnetograms observed by SDO/HMI, the twisted magnetic structure was found at the location of the filament (see Figs. 6c and 6d). After the formation of the first filament, it experienced a failed eruption and disappeared. The second filament formed quickly at the same location after the vanish of the first one. Interestingly, one foot-point of two active-region filaments is always rooted in the small rotating sunspot. According to these observations, Yan et al. (2015) suggested that shearing motion and small pore rotation play an important role in the formation of two active-region filaments.

Besides the case mentioned above, there is another case on the formation of an active region-filament reported by Yan et al. (2016)[33] and  summed up by Cheng et al. (2017)[34]. An inverse S-shaped filament formed in active region NOAA 11884 from 2013 October 31 to November 2. The clockwise rotation of a small positive sunspot around the main negative trailing sunspot dragged the chromospheric fibrils and resulted in the formation of a curved filament (see the dashed line 1 named CD in Figs. 7(a1)-7(a2)). Then the small positive sunspot cancelled with the negative magnetic flux to create a long inverse S-shaped active-region filament (see the dashed line named AD in Fig. 7(a3)). In the chromospheric H$\alpha$ observations, the formation of the inverse S-shaped active-region filament was due to the magnetic reconnection between the curved filament 1 (the dashed line CD) and the chromospheric fibrils 2 (the dashed line AB) (see Figs. 7(a1)-7(a2)). Using the non-linear force free field extrapolation (NLFFF) based on the vector magnetograms observed by SDO/HMI, the formation process of the magnetic structure supporting the filament was reproducted (see Figs. 7(b1)-7(b3)). In this case, the movement of the small sunspot and the magnetic cancellation play a key role in the formation of this filament.

Different from these cases mentioned above, Xue et al. (2017)[35] presented a case of an active-region filament formation due to the tether-cutting magnetic reconnection between two groups of chromospheric fibrils (see Fig. 8(a)). The two groups of the chromospheric fibrils L1 and L2 experienced tether-cutting reconnection at their joint and the active-region filament with the twisted structures formed (see Fig. 8(b)). Wang \& Liu (2019)[36] also mentioned this observational result in their review paper. The similar phenomenon were also reported by Yang et al. (2016)[37] and Chen et al. (2016)[38] by using SDO data.

In summary, several formation mechanisms of solar filaments are found based on NVST H$\alpha$ observations. These results are very important for us to understand the nature of solar filaments. 

\subsubsection{Magnetic structures of active-region filaments}
 How the magnetic structures support the cool plasma against gravity in solar filaments is an important issue in solar physics. 
 In order to explain the observations of the magnetic field configuration of solar filaments, there are two popular views proposed by some authors. One view is the sheared arcade model[39][40], and the other is the flux rope model[41][42][43][44]. 
 
Based on the NVST observations, several active-region filaments were observed during their formation and activation stages. Yang et al. (2014)[45] studied a filament activation by using H$\alpha$ data. They found that a filament was activated due to magnetic field cancellation. After the rise of the activated filament, the twisted flux rope structure was exhibited by tracking the flows along helical threads. Using the photospheric vector magnetograms, Yan et al. (2015) and Yan et al. (2016)[32][33] extrapolated the magnetic structures of two active-region filaments by using non-linear force free field model and found that there is a flux rope at the locations of the active-region filaments. According to the connections of the filament threads, Xue et al. (2016)[46] also found the threads of the filament winded each other and formed a flux rope. By tracking two rotating features during a filament eruption, Chen et al. (2019)[47] found that a total twist of about 1.3 $\pi$ is estimated to be stored in the filament before its eruption. These observations are inclined to support that active-region filaments may have twisted magnetic structures before their eruptions. Awasthi, Liu \& Wang (2019)[48] investigated the Doppler shifts and longitudinal oscillation in an active-region filament and found that the composite motions of the filament material suggest a double-decker host structure with mixed signs of helicity, comprising a flux rope atop a sheared-arcade system.

The exact magnetic structures of solar active-region filaments are still unclear. The measurement of chromospheric and coronal magnetic fields may help to solve this problem. 

\subsubsection{Filament eruption}
An eruptive filament is often taken as the bright core of a typical CME. The mechanism of filament eruption has been studied for many years. Several models have been proposed, e.g., kink instability[49][50], torus instability[51]), breakout model[52], tether-cutting model[53], catastrophic loss of equilibrium[54][55], and so on. Up to now, the trigger mechanism for a specified observed filament eruption remains unclear. 

By using NVST and SDO data, Wang et al. (2016)[56] reported that the opposite horizontal electric currents under the filament appeared before onset of the filament eruptions. They suggested that these opposite electric currents were carried by the new flux emerging from below the photosphere, which might be the trigger mechanism of the filament eruption. Figure 8 presents the observation of vector magnetograms and the electric current perpendicular to the axis of the filament. The yellow lines in Figs. 9(a) and 9(b) indicate the position of the cross-section of the electric current (Figs. 9(c) and 9(d)). Compared with Fig. 9(a) and Fig. 9(b), it is obvious that the opposite current appeared before the filament eruption. 

Sometimes, solar filaments consist of two parts, which have different twisted magnetic structures. Bi et al. (2015)[57] found a filament composed of two twisted flux ropes winding around each other. One has higher twist than the other. The one with higher twist erupted and the other with the lower twist part did not erupt. They suggested that the highly twisted flux rope became kink unstable when the instability threshold declined with the expansion of the flux rope. Cheng, Kliem, \& Ding (2018)[58] also found that some filaments just experienced partial eruption. Except for the individual filament eruption, two adjacent filaments can be observed to erupt in succession. This phenomenon is called sympathetic events. Li et al. (2017)[59] reported that the eruption of one of the two adjacent filaments is likely to be triggered by kink instability, while the weakening of overlying magnetic fields due to the magnetic reconnection at an X-point between the two filament systems might result in the onset of the other filament eruption. The two filament eruptions are confined due to the strong constraints of the overlying large-scale quiescent filament. Sometimes, the change of the overlying surrounding magnetic structure can also work as the trigger of the filament eruption. Zhou et al. (2017)[60] found that the rising of a flux rope resulted in magnetic reconnection between two external magnetic loops. The external magnetic reconnection weakens the constraining effect of the overlying field and leads to the further rising of the filament. Finally, the filament erupted due to torus instability. 

Yang \& Chen (2019)[61] presented state-of-the-art high-resolution H$\alpha$ and SDO/AIA multi-wavelength images tracking the evolution of an active region before, during, and after a blowout filament eruption that produced a CME. They found unambiguous observational evidences indicating that multiple interactions occurred between the emerging magnetic fields and the coronal loops overlying the filament. The interactions resulted in the formation of a sigmoid structure and the eruption of the filament. After the filament eruption, the emerging magnetic fields continued to reconnect with the remaining filament channel, leading to the reformation of the filament. In the photosphere, flux cancellation associated with $\delta$ sunspots formation occurred during the observation. These observations provide compelling evidence that tether-cutting reconnection accompanying magnetic flux cancellation first triggered the blowout eruption and then continued to restructure the magnetic field for the reformation of the filament. The observations demonstrate previously proposed mechanisms for the triggering of CME/flare eruptions and for the building of sheared fields where filaments form.

{\bf The aforementioned studies suggest that the trigger mechanism of solar filament eruptions may be diverse.} Prediction of solar filament eruptions is also a very interesting topic. 

\subsection{Observational evidence and the role of small-scale magnetic reconnection in the solar eruptions}
Magnetic reconnection is a very important process of energy release in Astrophysics. It is widely accepted that the free energy stored in the non-potential fields is released by magnetic reconnection during solar eruptions. There are several models of solar eruptions involving in magnetic reconnection. However, observational evidences have been relatively rare and typical only partial[62][63][64][65][66]. The role of magnetic reconnection in solar eruptions is still unclear.

\subsubsection{Observational evidence of small-scale magnetic reconnection}
Observational features of large-scale magnetic reconnection were found in some solar eruptions. However, small-scale magnetic reconnection in solar activities is rarely reported. Using the high spatial and temporal H$\alpha$ line center images observed by the NVST, the magnetic reconnection between two groups of chromospheric fibrils was discovered by Yang et al. (2015)[67] for the first time. L1 and L2 denote the two groups of the chromospheric fibrils  (see Fig. 10(a)). After the magnetic reconnection, the two newly formed groups of the chromospheric fibrils were labelled as L3 and L4 (see Fig. 10(b)). The thickness and length of the reconnection region are measured to be about 420 km and 1.4 Mm, respectively (see the blue box in Fig. 10(b)). Meanwhile, it is found that there were two stages of the magnetic reconnection. The first stage is a slow phase with a duration of more than several tens of minutes at the beginning of the magnetic reconnection. The second stage is a rapid step lasting for only about three minutes. The inflows along the slit AB in Fig. 10(a) range from 1.9 km $s^{-1}$ to 15.4 km $s^{-1}$ and the outflows range from 2.4 km $s^{-1}$ to 25.8 km $s^{-1}$. The down-flow along the newly formed cusp structures ranges from 40 km $s^{-1}$ to 80 km $s^{-1}$. 

Yang et al. (2016)[68] found another case of magnetic reconnection between two groups of loops with an X-shaped configuration in the chromosphere. They found that the anti-directed loops approached each other and then began to reconnect. New loops formed and stacked together. They also found that a set of loops suddenly retracted toward the balance position and performed an overshoot movement, leading to a convergent oscillation with the mean period of about 45 s. Due to the sudden retraction, other lower loops were pushed outward and performed an oscillation with the period of about 25 s.

\subsubsection{Release of twist in a filament by magnetic reconnection}
Except for the small-scale magnetic reconnection mentioned above, another magnetic reconnection event was also found during a filament eruption. Using high temporal and spatial resolution H$\alpha$ observations from the NVST, and supplemented by UV, EUV images and vector magnetograms observed by SDO, the X-ray data from Hinode and GOES, the magnetic reconnection occurring in the active region 12178 on October 03, 2014 was studied by Xue et al. (2016)[65] in detail. The magnetic reconnection occurs between the filament threads and the ambient chromospheric fibrils (see Figs.11(a) and 11(b)). For the first time, it was observed that the twist in the filament is released quickly by the magnetic reconnection. In this event, an unprecedented comprehensive set of observational evidences are detected, including reconnection inflows and outflows, current sheet, two hot cusp-shaped structures, newly formed magnetic loops, shrinking of the loops, and so on (see Fig. 11(c)-(f)). The estimated reconnection rate between 0.08 and 0.6 is consistent with the values of the theoretical expectations. In addition, the change of magnetic structures is supported by the change in the structure of the extrapolated nonlinear force-free field and by a data-constrained magnetohydrodynamic simulation (see Figs. 11(g)-11(i)). These results further confirm the occurrence of the magnetic reconnection during the filament eruption and the release process of the twist in the filament by the magnetic reconnection.

\subsubsection{Interchange magnetic reconnection between a filament and nearby open fields}
Interchange reconnection is defined as the interaction between closed and open fluxes[69]. It is considered to be a primary mechanism to transfer the dense mass, heat, and momentum to solar wind[70][71]. Though  this conception was proposed for many years, the observation evidences of interchange reconnection is very few (Kong et al. 2018)[72]. Using high resolution H$\alpha$ data from the NVST, Zheng et al. (2017)[73] studied a confined erupting filament and found evidence of interchange magnetic reconnection between the filament and adjacent open fields on 2014 December 24. During the filament eruption, the untwisting and rising motion of the filament bring their embedded closed field lines to meet with the adjacent oppositely open fields, and some of the closed field of the filament were opened. Moreover, the newly formed open field lines were observed to be rooted at negative polarities around one footpiont of the filament and the newly formed closed loops.

\subsubsection{Magnetic reconnection between a twisted arch filament system and coronal loops}

It is well accepted that magnetic reconnection plays an important role in the solar activities. Similar to the results of Xue et al. (2016)[65] and Zheng et al. (2017)[73], Huang et al. (2018)[74] found a reconnection event occurred between the threads of a twisted arch filament system (AFS) and coronal loops. Their observations reveal that the relaxation of the twisted AFS drives some of its threads to encounter the coronal loops, providing inflows of the reconnection. Magnetic reconnection can transfer the free magnetic energy in the twisted magnetic system into the upper solar atmosphere. 

\subsubsection{Magnetic reconnection between two active-region filaments}

Using high spatial and temporal resolution H$\alpha$ data from the NVST and simultaneous observations from the SDO, Yang et al. (2017)[75] investigated a rare event of the interaction between two filaments (F1 and F2) in AR NOAA 11967 on 2014 January 31 (see Fig. 12(a)). The adjacent two filaments were almost perpendicular to each other. Their interaction was driven by the movement of F1 and started when the two filaments collided with each other. During their interaction, the threads of F1 continuously slipped from the northeast to the southwest, and were accompanied by the brightenings at the junction of two filaments and the northeast footpoint of F2 (Figs. 12(b)-12(c)). At the same time, bright material initiated from the junction of two filaments was observed to move along F1. The magnetic connectivities of F1 were found to change after their interaction (see Fig. 12(d)). These observations suggest that magnetic reconnection was involved in the interaction of two filaments and resulted in the eruption of F2.

\subsubsection{Oscillatory magnetic reconnection}

Xue et al. (2019)[76] found that a small-scale oscillatory magnetic reconnection occurred in active region 11800 from July 24 to 25, 2013. It consists of four relatively independent magnetic reconnections which last for about 48, 158, 275, and 340 minutes, respectively. Moreover, the inflow/outflow regions of the previous magnetic reconnection are transformed into the outflow/inflow regions of the following magnetic reconnection, and a current sheet was observed in each magnetic reconnection. Four current sheets along two nearly perpendicular directions are formed alternately. Therefore, the oscillatory reconnection experiences two cycles, and the periods of the two oscillations are 206 and 615 minutes, respectively, which are much longer than those in the previous results. In addition, a small-scale flux rope is observed to form and disappear, which is caused by the oscillatory magnetic reconnection.

More and more evidences of magnetic reconnection are provided by the NVST observation. It is a fact that magnetic reconnection plays an important role in solar eruptions[55][77]. However, how the particles are accelerated in the current sheet is still obscure. The simulations showed that there are many magnetic islands in current sheet[78][79][80]. So far, there is no reliable observational evidence that magnetic islands exist in current sheets. Higher resolution observational data are required to reveal this phenomenon.

\subsection{Small-scale eruptive activities}
Following the improvement of the observation technology, many small-scale eruptive activities in the solar atmosphere are observed clearly. For example, there are various kinds of jet-like structures like spicules, chromospheric jets, X-ray/UV/EUV jets, etc, and sudden brightening in the extended H$\alpha$ wings named Ellerman bombs in the different solar atmosphere[81]. Studying the triggering mechanisms of these phenomena is one of the important subjects in solar physics.

\subsubsection{Ellerman bombs}
Ellerman bombs (EBs)[81] are defined as the two emission bumps in both wings of chromospheric lines, such as H$\alpha$ 6563 \AA\ and Ca II 8542 \AA\ lines. They are small-scale brightening events in the solar lower atmosphere. Recent high-resolution observations by the Interface Region Imaging Spectrograph (IRIS)[82] revealed that absorption lines (singly ionized or neutral) superimposed on the greatly broadened transition region lines[83] are typically found in small-scale compact bright regions from slit-jaw images (SJIs) taken with the 1400 \AA\ and 1330 \AA\ filters. This phenomenon indicates that there is local heating of the photosphere or lower chromosphere to 8 $\times$ $10^{4}$ K under the assumption of collisional ionization equilibrium named UV bursts. Is there a connection between UV bursts and EBs? Tian et al. (2016)[84] identified 10 UV bursts by using the IRIS observations. They found that three UV bursts are unambiguously and three UV bursts are possibly connected to EBs, which show intense brightening in the extended H$\alpha$ wings without leaving an obvious signature in the H$\alpha$ core (see Fig. 13). Their findings support the formation of these bombs in the photosphere and EBs can be heated much more efficiently than previously thought. They also found that the Mg II k and h lines can be used to identify EBs similarly to H$\alpha$, which opens a promising new window for EB studies. The remaining four UV bursts have no connection to EBs. These UV bursts were suggested to form in a higher layer, possibly in the chromosphere.

Chen et al.(2017)[85] have carried out a statistical study on the visibility of EBs in the AIA 1700 \AA\ images by using NVST $H\alpha$ core and off-band images. There are 145 EBs that were identified from $H\alpha$ wing images, in which 74 EBs (about 51\%) can be clearly identified as bright points (BPs) in the AIA 1700 \AA\ images. On the contrary, they have identified 125 BPs in the 1700 \AA\ images, in which 66 BPs (53\%) corresponds to EBs in the H$\alpha$ wing images. These results show that most of the small-scale, compact, and transient brightenings in the AIA 1700 \AA\ images can be identified as EBs.

Up to now, the heating mechanism of EBs is  still controversial. Based on the connection between UV bursts and EBs, Tian et al. (2016) concluded that some EBs could be heated to 1-8 $\times$$10^{4}$, which is much higher than previously thought. To address this issue, Ni et al. (2016)[86] carried out MHD simulations and found that reconnection around the temperature minimum region (TMR) can indeed heat some of the materials to such high temperature if the plasma beta is low. Fang et al. (2017)[87] have performed non-LTE calculations H$\alpha$ and Ca II 8542 \AA\ line profile and found that the heating of EBs cannot reach 10000 K. Consequently, the observations and theoretical studies are not consistent with each other. The heating of EBs is an important problem for the understanding of magnetic reconnection in the partially ionized lower solar atmosphere. In the future, observations and theoretical studies are required.

\subsubsection{Mini-filaments and jets}
Due to the high resolution H$\alpha$ observation, Hong et al. (2017)[88] studied a mini-filament eruption in active region NOAA 12259 on 2015 January 14. The mini-filament experienced twice eruptions (a partial and a full eruption). The first eruption was associated with a fan-spine jet and the second eruption was associated with a blowout jet. It is worth mentioning that the blowout jet is accompanied by an interplanetary type-III radio burst. This is the first time to report that the mini-filament eruption is correlated with the interplanetary type-III radio burst. Tian et al. (2017)[89] reported two successive two-sided loop jets caused by magnetic reconnection between two adjacent filamentary threads, which were observed by the NVST at the southern periphery of active region NOAA 12035 on 2014 April 16. 

Previous modeling and observations of jets support the occurrence of jets due to magnetic reconnection between emerging bi-polar and their overlying horizontal magnetic fields. The observations of Tian et al. (2017)[89] extend the jet model, i.e., a ``tether-cutting model", which is often used to explain the formation of large-scale filaments or flux ropes. Similar to the observations of Hong et al. (2017)[88], Shen et al. (2017)[90] also studied a blowout jet that experienced two distinct ejection stages at the southern periphery region of active region NOAA 12035 on 2014 April 16 (see Fig. 14). In the first stage, a small jet is triggered by the magnetic reconnection between the rising confining loops and the overlying open fields. A blowout jet is triggered by magnetic reconnection between the rising confining loops and the overlying open field in the second stage. They also observed  periodic metric radio type III bursts at the every beginning of the two stages. In addition, two-side-loop jets were found by Zheng et al. (2018)[91] by using the observation of the NVST in active region NOAA 12681 on 2017 September 23. These jets were triggered by magnetic reconnection between the emerging loops and the overlying horizontal filament threads.

Using the data from SDO and NVST, a circular filament eruption associated with the formation of jet under a non-axisymmetric fan spine configuration was studied. It is found that the non-axisymmetric fan spine configuration can lead to a non-radial eruption of filament toward magnetic null point. In this process, the eruptive filament interacted with the overlying magnetic loops, resulting in the disintegration of the filament, which caused the formation of a blowout jet[92].

\subsection{Formation and trigger mechanism of solar flares}

The flares associated without CMEs are termed ``confined flares"[93]. Using the H$\alpha$ observations from the NVST, Yang et al. (2014)[94] investigated the fine structures of three confined flares on 2013 October 12 (see Fig. 15). All the three confined flares took place successively at the same location in active region NOAA 11861. They have similar morphologies and are termed as homologous confined flares. Supplemented by the SDO observations, many large-scale coronal loops above the confined flares are clearly found in multi-wavelength observations. A reconnection between the H$\alpha$ fibrils of two emerging dipoles resulted in the change of the H$\alpha$ fibril configuration. The reconnection also occurs between a set of emerging H$\alpha$ fibrils and a set of pre-existing large-scale loops, forming a typical three-legged structure. Liu et al. (2018)[95] also found that a C-class flare was triggered by magnetic reconnection between an emerging magnetic bi-pole region and the super-penumbral fibrils. These observations confirmed that the magnetic reconnection between the emerging loops and the pre-existing loops triggers solar flares and the large-scale overlying loops prevent the flares from erupting. 

Based on multi-wavelength observations from the NVST and SDO, Xu et al. (2019)[96] found that the sunspot structures and their magnetic fields suffered significant transformations during an M-class flare. One penumbral segment decayed with partial fields transforming into umbra fields when swept by one flare ribbon. At the same time, an adjacent penumbral segment expanded permeating the granular area with the horizontal fields enhancing along the flaring PIL. Their observations support the magnetic implosion idea that the flare eruptions in the upper atmosphere could result in the magnetic field rearrangement on the surface of the Sun.

In recent years, circular-ribbon flares have caught a lot of attentions due to challenging the traditional model of two-ribbon flares. Using the NVST H$\alpha$ data, Li et al. (2018)[97] studied a mini-filament erupting in association with a circular ribbon flare on 2014 March 17. The circular flare ribbon was observed around the filament at the onset of the eruption. After the filament activation, its eruption looked like the formation of a surge, which ejected along one end of a large-scale closed coronal loops with a curtain-like shape. They carried out a potential field extrapolation and found that a null point was located above the mini-filament. They suggested that the null point reconnection may facilitate the eruption of the filament to produce the circular-ribbon flare. Xu et al. (2017)[98] reported two homologous circular-ribbon flares associated with two filament eruptions observed by the NVST in active region NOAA 11991 on 2014 March 5. Two small-scale filaments erupted in sequence associated with two homologous circular-ribbon flares and displayed an apparent writhing motion. Moreover, the post-flare loops and subsequent helical structures are clearly seen during their eruption. These observations imply that the small-scale filaments have twisted magnetic structures. In addition, Song \& Tian (2018a,b)[99][100] found that the eruptions of the mini-filaments are often associated with white-light flares. 

Recent investigations reveal that the active regions with rotating sunspots have relatively high solar flare productivity[101]. For example, active regions, e.g., NOAA 10486, NOAA 10930, NOAA 11158, produced many flares and coronal mass ejections (CMEs). However, the relationship between evolution of rotating sunspots and solar activities is still unclear.  

The triggering process of solar flares is not very clear. In order to address this issue, Yan et al. (2018)[102] used HMI data to study the occurrence of two successive X-class flares and two coronal mass ejections (CMEs) triggered by shearing motion and sunspot rotation in active region NOAA 12673 on 2017 September 6. From the evolution in the continuum intensity images and the line-of-sight magnetograms, it is very obvious that there was a shearing motion between the two sunspots from September 5 to the onset of the second X-class flare on September 6. Moreover, the sunspot with negative polarity rotated around its umbral center and the other sunspot with positive polarity also exhibited a slow rotation (see Fig.16(a) and 16(b)). The structure of the sunspot with negative polarity is very complex seen from high resolution TiO images in Figs. 16(c) and 16(d). Before the occurrence of the first X-class flare, it is found that there is a flux rope that forms along the polarity inversion line between the two sunspots (see Fig. 17). The S-shaped structure seen from the EUV observation corresponds to this flux rope. Moreover, the first X-class flare is associated with the eruption of the flux rope. 

The sunspot with negative polarity at the northwest of active region also began to rotate counter-clockwise before the onset of the first X-class flare, which is related to the formation of the second S-shaped structure in the EUV observation. The eruption of the second S-shaped structure produced the second X-class flare and a CME. The successive formation and eruption of two S-shaped structures were closely related to the counter-clockwise rotation of three sunspots. The existence of a flux rope is found prior to the onset of two flares by using non-linear force free field extrapolation based on the vector magnetograms observed by SDO/HMI. These results suggest that shearing motion and sunspot rotation play an important role in the buildup of the free energy and the formation of flux ropes in the corona.

\section{Summary and prospect}
The NVST is a very powerful solar telescope to observe the photosphere and chromosphere of the Sun. High spatial and temporal observational data from the NVST are very suitable for researchers to study small-scale solar activities. Based on the these data, many new results were obtained. 

By using TiO images, the photospheric BPs were studied in detail. The area, equivalent diameter, lifetime, horizontal velocity, diffusion index, motion range, and motion type of BPs were determined by using a new method named the Laplacian and morphological dilation algorithm[11][103]. Moreover, the relationships between the magnetic field intensity and the parameters of BPs were investigated. The diameters, intensities, lifetimes and velocities of UDs were found to depend on the surrounding magnetic field. In addition to the typical 5-minute and 3-minute oscillations in the photosphere and chromosphere of the Sun, one minute oscillation at different heights above sunspot umbra was discovered. 

Due to the high resolution chromospheric data (H$\alpha$ line center and off-band images), the fine structures and formation mechanisms of solar prominences or filaments were studied in detail. In addition to the counter-streaming flows, the kelvin-Helmholtz instability was identified in quiescent prominences. Moreover, the parallel tube-shaped structures in the barb were found during the evolution of the prominence. Magnetic reconnection between two groups of chromospheric fibrils driven by sunspot movement, shearing motion and sunspot rotation, and tether-cutting magnetic reconnection were found to be the formation mechanisms of solar active-region filaments. Note that there are two mechanisms on the filament channel formation, i.e., surface effects that reconfigure pre-existing coronal fields and magnetic emergence of a horizontal flux rope (see the review of Mackay et al. (2010)[104]. The results on the filament formation obtained using NVST data belong to the former mechanism. Though many works reported that solar eruptions are closely related to the shearing motion and sunspot rotation[105][106][107], the filament formation related to a small rotating pore was reported for the first time. 

The observational evidences of small-scale magnetic reconnection in solar eruptions are confirmed. Small-scale magnetic reconnection between the two groups of the small chromospheric fibrils and the release of twist in a filament by magnetic reconnection were discovered for the first time. Interchange magnetic reconnection between a filament and nearby open fields, magnetic reconnection between a twisted arch filament system and coronal loops, oscillatory magnetic reconnection, and magnetic reconnection between two active-region filaments were observed. The relatively large field-of-view of the NVST provides the opportunity to observe these different types of magnetic reconnection between the different magnetic structures. In comparison to the previous studies, all of the magnetic reconnection events observed by the NVST are on-disc and small-scale events. 

Through analyzing the relationship between UV bursts and EBs, the results are inclined to support that the formation of EBs was in the photosphere and EBs can be heated much more efficiently than previously thought. Moreover, the UV bursts were suggested to form in a higher layer, possibly in the chromosphere. 

The physical connections between the eruptions of mini-filaments and solar flares/jets were studied in detail. Using NVST data, the mini-filament eruption correlated with the interplanetary type-III radio burst was reported for the first time. Besides, two successive two-sided loop jets caused by magnetic reconnection between two adjacent filamentary threads were observed.

By using H$\alpha$ line center images, a series of confined flares were studied in detail. It is evidenced that magnetic reconnection between two groups of chromospheric fibrils led to the release of free energy in a series of successive confined flares. Part of magnetic fields in the penumbral segment transforming into umbra fields were found during another M-class confined flare. This result supports that the eruptions in the upper atmosphere can result in the magnetic field rearrangement in the lower atmosphere.

In the future, magnetograms with high spatial and temporal resolution will be provided at Fuxian lake observatory. High resolution magnetograms will be taken to study the magnetic field evolution of small-scale features in the photosphere such as BPs, foot-points of mini-filaments, EBs, and so on. Moreover, the spectrometers will start routine observations. According to the requirement of researchers, several observational models will be set up for users to choose. Note that these works are in progress. It is expected that more new results can be obtained due to the high resolution photospheric and chromospheric observations in combination with high resolution photospheric magnetograms.

References

[1] Liu Z , Xu J , Gu B Z , et al. New vacuum solar telescope and observations with high resolution. Research in Astronomy and Astrophysics, 2014, 14:705-718

[2] Xu Z , Jin Z Y , Xu F Y , et al. Primary observations of solar filaments using the multi-channel imaging system of the New Vacuum Solar Telescope. Proceedings of the International Astronomical Union, 2014, 8:117-120

[3] Wang R, Xu Z, Jin Z Y, et al. The first observation and data reduction of the Multi-wavelength Spectrometer on the New Vacuum Solar Telescope. Research in Astronomy and Astrophysics, 2013, 13:1240-1254

[4] Cai Y F , Xu Z , Chen Y C , et al. Composing Method for the Two-dimensional Scanning Spectra Observed by the New Vacuum Solar Telescope. Research in Astronomy and Astrophysics, 2018, 18:42-52

[5] Xiang Y Y , Liu Z , Jin Z Y . High resolution reconstruction of solar prominence images observed by the New Vacuum Solar Telescope. New Astronomy, 2016, 49:8-12

[6] Weigelt G P . Modified astronomical speckle interferometry ¡°speckle masking¡±. Optics Communications, 1977, 21:55-59

[7] Liu Z , Qui Y , Ke L , et al. New progress in the high resolution speckle imaging at Yunnan Observatory. Acta Astronomica Sinica, 1998, 39:217-224

[8] Zhu X , Wang H , Du Z , et al. Forced field extrapolation of the magnetic structure of the Halpha fibrils in solar chromosphere. The Astrophysical Journal, 2016, 826:51-58

[9] Cai Y , Xu Z , Li Z , et al. Precise Reduction of Solar Spectra Observed by the One-Meter New Vacuum Solar Telescope. Solar Physics, 2017, 292:150-163

[10] Ji K F , Xiong J P , Xiang Y Y , et al. Investigation of intergranular bright points from the New Vacuum Solar Telescope. Research In Astronomy And Astrophysics, 2016, 16:69-80

[11] Feng S , Deng L , Yang Y , et al. Statistical study of photospheric bright points in an active region and quiet Sun. Astrophysics And Space Science, 2013, 348:17-24

[12] Liu Y, Xiang Y, Erd{\'e}lyi R, et al. Studies of Isolated and Non-isolated Photospheric Bright Points in an Active Region Observed by the New Vacuum Solar Telescope. The Astrophysical Journal, 2018, 856: 17-34

[13] Yang, Y.-F., Lin, J.-B., Feng, S., et al. Evolution of isolated G-band bright points: size, intensity and velocity, Research in Astronomy and Astrophysics, 2014, 14:741-752

[14] Ji H S, Cao W D, Goode Philip R. Observation of Ultrafine Channels of Solar Corona Heating. The Astrophysical Journal Letters, 2012, 750:L25-L30

[15] Ji K F, Jiang X, Feng S, et al. Investigation of Umbral Dots with the New Vacuum Solar Telescope. Solar Physics, 2016, 291:357-69

[16] Feng S, Xu Z, Wang F, et al. Automated Detection of Low-Contrast Solar Features Using the Phase-Congruency Algorithm. Solar Physics, 2014, 289:3985-3994

[17] Su J T , Ji K F , Cao W , et al. Observations Of Oppositely Directed Umbral Wavefronts Rotating In Sunspots Obtained From The New Solar Telescope Of BBSO. The Astrophysical Journal, 2016, 817:117-133

[18] Su J T , Ji K F , Banerjee D , et al. Interference Of The Running Waves At Light Bridges Of A Sunspot. The Astrophysical Journal, 2016, 816:30-41

[19] Wang F , Deng H , Li B , et al. High-frequency Oscillations in the Atmosphere above a Sunspot Umbra. The Astrophysical Journal Letters, 2018, 856:L16-L34

[20]Zhou X , Liang H . The relationship between the 5-min oscillation and 3-min oscillations at the umbral/penumbral sunspot boundary. Astrophysics and Space Science, 2017, 362:46-56

[21] Zhou X , Liang H F , Li Q H , et al. Statistical research of the umbral and penumbral oscillations. New Astronomy, 2017, 51:86-95

[22] Yang S , Zhang J , Jiang F , et al. oscillating light wall above a sunspot light bridge. The Astrophysical Journal Letters, 2015, 804:L27-L32

[23] Hou Y J , Li T , Yang S H , et al. Light Walls Around Sunspots Observed by the Interface Region Imaging Spectrograph. Astronomy \& Astrophysics, 2016, 589:L7-L11

[24] Tian H , Yurchyshyn V , Peter H , et al. Frequently Occurring Reconnection Jets from Sunspot Light Bridges. The Astrophysical Journal, 2018, 854:92-105

[25] Forbes T G . A review on the genesis of coronal mass ejections. Journal of Geophysical Research Space Physics, 2000, 105:23153-23166

[26] Martin S F . Conditions for the Formation and Maintenance of Filaments šC (Invited Review). Solar Physics, 1998, 182:107-137

[27] Yan X L , Xue Z K , Xiang Y Y , et al. Fine-scale structures and material flows of quiescent filaments observed by New Vacuum Solar Telescope. Research in Astronomy and Astrophysics, 2015, 15:1725-1734

[28]Li H , Liu Y , Tam K V , et al. Piecewise mass flows within a solar prominence observed by the New Vacuum Solar Telescope[J]. Astrophysics and Space Science, 2018, 363:118-123

[29] Shen Y D , Liu Y , Liu Y D , et al. Fine Magnetic Structure and Origin of Counter-streaming Mass Flows in a Quiescent Solar Prominence. The Astrophysical Journal Letters, 2015, 814: L17-L26

[30] Yang H, Xu Z, Lim E K, et al.Observation of the Kelvin-Helmholtz Instability in a Solar Prominence. The Astrophysical Journal, 2018, 857:115-124

[31] Li D , Shen Y D , Ning Z J , Zhang Q M , et al. Two Kinds of Dynamic Behavior in a Quiescent Prominence Observed by the NVST. The Astrophysical Journal, 2018, 863:192-203

[32] Yan X L , Xue Z K , Pan G M , et al. The Formation and Magnetic Structures of Active-Region Filaments Observed By NVST, SDO, and Hinode. The Astrophysical Journal Supplement Series, 2015, 219:17-36

[33] Yan X L, Priest E R , Guo Q L , et al. The Rapid Formation of a Filament Caused by Magnetic Reconnection between Two Sets of Dark Threadlike Structures. Astrophysical Journal, 2016, 832:23-35

[34] Cheng X, Guo Y,  Ding M. Origin and structures of solar eruptions II: Magnetic modeling. Science in China Earth Sciences, 2017,60:1408-1439

[35] Xue Z K, Yan X L, Yang L H, et al. Observing Formation of Flux Rope by Tether-cutting Reconnection in the Sun. The Astrophysical Journal Letters, 2017, 840:L23-L31

[36] Wang,H., Liu,C.\ 2019, Frontiers in Astronomy and Space Sciences, 6, 00018

[37] Yang B , Jiang Y , Yang J , et al. The Rapid Formation Of A Filament Caused By Magnetic Reconnection Between Two Sets Of Dark Threadlike Structures. The Astrophysical Journal, 2016, 816:41-50

[38] Chen H , Zhang J , Li L , et al. Tether-cutting Reconnection between Two Solar Filaments Triggering Outflows and a Coronal Mass Ejection. The Astrophysical Journal Letters, 2016, 818:27-34

[39]Antiochos S K , Dahlburg R B , Klimchuk J A . The Magnetic Field of Solar Prominences. The Astrophysical Journal Letters, 1994, 420:L41-L45

[40]Devore C R , Antiochos S K . Dynamical Formation and Stability of Helical Prominence Magnetic Fields. Astrophysical Journal, 2000, 539:954-963

[41]Demoulin P , Priest E R . A twisted flux model for solar prominences. II - Formation of a dip in a magnetic structure before the formation of a solar prominence. Astronomy and Astrophysics, 1989, 214:360-368

[42]Amari T , Luciani J F , Mikic Z , et al. A Twisted Flux Rope Model for Coronal Mass Ejections and Two-Ribbon Flares. The Astrophysical Journal Letters, 2000, 529:L49-L52

[43] Jiang C W, Wu S T, Feng X S, Hu Q. Nonlinear Force-free Field Extrapolation of a Coronal Magnetic Flux Rope Supporting a Large-scale Solar Filament from a Photospheric Vector Magnetogram. The Astrophysical Journal Letters, 2014, 786:L16-L23

[44]Su Y , van Ballegooijen A, McCauley P, et al. Magnetic Structure and Dynamics of the Erupting Solar Polar Crown Prominence on 2012 March 12.The Astrophysical Journal, 2015, 807:144-161

[45]Yang S , Zhang J , Liu Z , et al. New Vacuum Solar Telescope Observations Of A Flux Rope Tracked By A Filament Activation. The Astrophysical Journal Letters, 2014, 784:L36-L43

[46] Xue Z K , Yan X L , Zhao L , et al. Failed eruptions of two intertwining small-scale filaments.Publications of the Astronomical Society of Japan, 2016, 68:7-19

[47]Chen H , Zheng R , Li L , et al. Untwisting and Disintegration of a Solar Filament Associated with Photospheric Flux Cancellation. The Astrophysical Journal, 2019, 871:229-240

[48]Awasthi A K , Liu R , Wang Y. Double-Decker Filament Configuration Revealed by Mass Motions. The Astrophysical Journal, 2019, 872:109-120

[49]Hood A W , Priest E R . Kink instability of solar coronal loops as the cause of solar flares. Solar Physics, 1979, 64:303-321

[50]T{\"o}r{\"o}k T, \& Kliem B. Confined and Ejective Eruptions of Kink-unstable Flux Ropes. The Astrophysical Journal, 2005, 630:L97-L100

[51]Kliem B , T{\"o}r{\"o}k T. Torus Instability. Physical Review Letters, 2006, 96:255002.

[52] Antiochos S K , Devore C R , Klimchuk J A . A Model for Solar Coronal Mass Ejections. The Astrophysical Journal, 1999, 510:485-493

[53]Moore R L , Sterling A C , Hudson H S , et al. Onset of the Magnetic Explosion in Solar Flares and Coronal Mass Ejections. The Astrophysical Journal, 2001, 552:833-848

[54]Forbes T G , Isenberg P A . A catastrophe mechanism for coronal mass ejections. The Astrophysical Journal, 1991, 373:294-307

[55]Lin J , Forbes T G . Effects of reconnection on the coronal mass ejection process. Journal of Geophysical Research, 2000, 105:2375-2392

[56] Wang J C, Yan X L, Qu Z Q , et al. The Evolution of the Electric Current during the Formation and Eruption of Active-region Filaments The Astrophysical Journal, 2016, 817:156-166

[57]Bi Y, Jiang Y C, Yang J Y, et al. Partial Eruption of a Filament with Twisting Non-uniform Fields.The Astrophysical Journal, 2015, 805:48-56

[58]Cheng X, Kliem B, Ding M D. Unambiguous Evidence of Filament Splitting-Induced Partial Eruptions. The Astrophysical Journal, 2018, 856:48-63

[59] Li S ,Su Y , Zhou T , et al. High-resolution Observations of Sympathetic Filament Eruptions by NVST. The Astrophysical Journal, 2017, 844:70-83

[60] Zhou G P , Zhang J , Wang J X , et al. A Study of External Magnetic Reconnection that Triggers a Solar Eruption. The Astrophysical Journal Letters, 2017, 851:L1-L9

[61]Yang B, Chen H. Filament Eruption and Its Reformation Caused by Emerging Magnetic Flux. The Astrophysical Journal, 2019, 874:96-110

[62] Zhang J , Yang S , Li T , et al. Magnetic Reconnection: From "Open" Extreme-ultraviolet Loops to Closed Post-flare Ones Observed by SDO. The Astrophysical Journal, 2013, 776:57-66

[63] Su Y , Veronig A M , Holman G D , et al. Imaging coronal magnetic-field reconnection in a solar flare. Nature Physics, 2013, 9:489-493

[64]Tian H, Li G, Reeves K K, et al. Imaging and Spectroscopic Observations of Magnetic Reconnection and Chromospheric Evaporation in a Solar Flare. The Astrophysical Journal Letters, 2014, 797:L14-L21

[65]Xue Z , Yan X , Cheng X , et al. Observing the release of twist by magnetic reconnection in a solar filament eruption. Nature Communications, 2016, 7:11837

[66]Li L , Zhang J , Peter H , et al. Magnetic reconnection between a solar filament and nearby coronal loops. Nature Physics, 2016, 12:847-851

[67]Yang S, Zhang J, Xiang Y. Magnetic reconnection between small-scale loops observed with the New Vacuum Solar Telescope.The Astrophysical Journal Letters, 2015, 798:L11-L17

[68]Yang S , Xiang Y . Oscillation of Newly Formed Loops after Magnetic Reconnection in the Solar Chromosphere. The Astrophysical Journal Letters, 2016, 819:L24-L30

[69]Crooker N U , Gosling J T , Kahler S W . Reducing heliospheric magnetic flux from coronal mass ejections without disconnection. Journal of Geophysical Research (Space Physics), 2002, 107:1028-1033

[70] Wang Y M , Sheeley , N R , Walters J H , et al. Origin of Streamer Material in the Outer Corona. The Astrophysical Journal, 1998, 498:L165-L168

[71] Fisk L A , Schwadron N A , Zurbuchen T H . Acceleration of the fast solar wind by the emergence of new magnetic flux. Journal of Geophysical Research, 1999, 104:19765-19772

[72]Kong D F , Pan G M , Yan X L , et al. Observational Evidence of Interchange Reconnection between a Solar Coronal Hole and a Small Emerging Active Region. The Astrophysical Journal Letters, 2018, 863:L22-L29

[73]Zheng R , Chen Y , Wang B , et al. Interchange Reconnection Associated with a Confined Filament Eruption: Implications for the Source of Transient Cold-dense Plasma in Solar Winds. The Astrophysical Journal, 2017, 840:3-12

[74]Huang Z, Mou C, Fu H, et al. A magnetic reconnection event in the solar atmosphere driven by relaxation of a twisted arch filament system. The Astrophysical Journal Letters, 2018, 853:L26-l34

[75]Yang L H, Yan X L, Li T , et al. Interaction of Two Active Region Filaments Observed by NVST and SDO. The Astrophysical Journal, 2017, 838:131-139

[76]Xue Z K, Yan X L, Jin C L, et al. A Small-scale Oscillatory Reconnection and the Associated Formation and Disappearance of a Solar Flux Rope. The Astrophysical Journal Letters, 2019,874:L27-L36

[77]Mei Z X , Keppens R , Roussev I I , et al. Parametric study on kink instabilities of twisted magnetic flux ropes in the solar atmosphere. Astronomy and Astrophysics, 2018, 609:A2-A15

[78]Ni L, Kliem B, Lin J, et al. Fast magnetic reconnection in the solar chromosphere mediated by the plasmoid instability. The Astrophysical Journal, 2015, 799:79-95

[79]Mei Z X , Keppens R , Roussev I I , et al. Magnetic reconnection during eruptive magnetic flux ropes. Astronomy and Astrophysics, 2017, 604:L7-L11

[80]Ye J , Lin J , Raymond J C , et al. Numerical Study of the Cascading Energy Conversion of the Reconnecting Current Sheet in Solar Eruptions.Monthly Notices of the Royal Astronomical Society, 2017, 482:588-605

[81]Ellerman F . Solar hydrogen "bombs". Astrophysical Journal, 1917, 46:298-302

[82]De Pontieu B, Title A M, Lemen J R, et al. The Interface Region Imaging Spectrograph (IRIS).Solar Physics, 2014, 289:2733-2779

[83]Peter H , Tian H , Curdt W , et al. Hot explosions in the cool atmosphere of the Sun. Science, 2014, 346:1255726

[84]Tian H , Xu Z , He J , et al. ARE IRIS BOMBS CONNECTED TO ELLERMAN BOMBS?. The Astrophysical Journal, 2016, 824:96-110

[85] Chen,Y.J.,Tian,H., Xu,Z.,Xiang,Y.Y., Fang,Y.L.,Yang,Z.H. Ellerman bombs observed with the new vacuum solar telescope and the atmospheric imaging assembly onboard the solar dynamics observatory, Geoscience Letters, 2017,4:30-35

[86]Ni L , Lin J , Roussev I I , et al. Heating Mechanisms in the Low Solar Atmosphere through Magnetic Reconnection in Current Sheets. The Astrophysical Journal, 2016, 832:195-206

[87]Fang C , Hao Q , Ding M D , \& Li Z. Can the temperature of Ellerman Bombs be more than 10 000 K?. Research in Astronomy and Astrophysics, 2017, 17:31-37

[88]Hong J , Jiang Y , Yang J , et al. Minifilament Eruption as the Source of a Blowout Jet, C-class Flare, and Type-III Radio Burst. The Astrophysical Journal, 2017, 835:35-46

[89]Tian Z , Liu Y , Shen Y , et al. Successive Two-sided Loop Jets Caused by Magnetic Reconnection between Two Adjacent Filamentary Threads. The Astrophysical Journal, 2017, 845:94-102

[90]Shen Y D, Liu Y D , Su J T , et al. On a solar blowout jet: driven mechanism and the formation of cool and hot components. The Astrophysical Journal, 2017, 851:67-80

[91]Zheng R, Yao C, Huang Z, et al. Two-sided-loop Jets Associated with Magnetic Reconnection between Emerging Loops and Twisted Filament Threads. The Astrophysical Journal, 2018, 861:108-119

[92]Li H , Yang J . A Fan Spine Jet: Nonradial Filament Eruption and the Plasmoid Formation. The Astrophysical Journal, 2019, 872:87-96

[93]Svestka Z , Jackson B V , Machado M E . Eruptive Solar Flares. Lecture Notes in Physics, 1992, 399:1

[94]Yang S , Zhang J , Xiang Y . Fine Structures and Overlying Loops of Confined Solar Flares. Astrophysical Journal Letters, 2014, 793:L28-L35

[95]Liu S, Zhang H Q , Choudhary D P , Srivastava A K , \& Nath Dwivedi B. Superpenumbral chromospheric flare. Research in Astronomy and Astrophysics, 2018, 18:130-140

[96]Xu Z, Yang J Y, Ji K F, Bi Y, \& Yang B. Magnetic Field Rearrangement in the Photosphere Driven by an M5.0 Solar Flare.The Astrophysical Journal, 2019, 874:134-143

[97] Li, H., Yang, J., Jiang, Y., Bi, Y., Qu, Z., and Chen, H. The surge-like eruption of a miniature filament associated with circular flare ribbon, Astrophysics and Space Science, 2018, 363: 26-31.

[98]Xu Z , Yang K , Guo Y , et al. Homologous Circular-ribbon Flares Driven by Twisted Flux Emergence. The Astrophysical Journal, 2017, 851:30-41

[99]Song Y L , \& Tian H. Investigation of White-light Emission in Circular-ribbon Flares. The Astrophysical Journal, 2018, 867:159-172

[100] Song, Y.~L., Guo, Y., Tian, H., Zhu, X.~S., Zhang, M., Zhu, Y.~J. Observations of a White-light Flare Associated with a Filament Eruption.\ The Astrophysical Journal, 2018, 854, 64-86.

[101] Yan X L , Qu Z Q , Kong D F . Relationship between rotating sunspots and flare productivity. Monthly Notices of the Royal Astronomical Society, 2008, 391:1887-1892

[102] Yan X L , Wang J C , Pan G M , et al. Successive X-class Flares and Coronal Mass Ejections Driven by Shearing Motion and Sunspot Rotation in Active Region NOAA 12673. The Astrophysical Journal, 2018, 856:79-93

[103]Yang Y , Li Q , Ji K , et al. On the Relationship Between G-Band Bright Point Dynamics and Their Magnetic Field Strengths. Solar Physics, 2016, 291:1089-1105

[104]Mackay D H , Karpen J T , Ballester J L , et al. Physics of Solar Prominences: II. Magnetic Structure and Dynamics. Space Science Reviews, 2010, 151:333-399

[105] Yan X L, Qu Z Q. Rapid rotation of a sunspot associated with flares. Astronomy \& Astrophysics, 2007, 468:1083-1088

[106] Yan X l, Qu Z Q, Xu C L, et al. The causality between the rapid rotation of a sunspot and an X3.4 flare. Research in Astronomy and Astrophysics, 2009, 9:596-602.

[107] Su Y , Golub L , Van Ballegooijen A , et al. Evolution of the Sheared Magnetic Fields of Two X-Class Flares Observed by Hinode/XRT. Publications of the Astronomical Society of Japan, 2007, 5:S785-S791

\noindent{\bf Acknowledgements}\\
\noindent We would like to thank the NVST, SDO/AIA, SDO/HMI teams for high-cadence data support. This work is sponsored by the National Science Foundation of China (NSFC) under the grant numbers (11873087, 11633008), by the Youth Innovation Promotion Association CAS (No.2011056), by the Yunnan Talent Science Foundation of China (2018FA001), by Project Supported by the Specialized Research Fund for State Key Laboratories, and by the grant associated with project of the Group for Innovation of Yunnan Province.

\begin{table}\tiny
\begin{center}
\caption[]{Properties of the BPs of the Six Data Sets. Table reproduced from Ji et al. (2016), copyright by RAA}\label{Tab:tab2}
  \begin{tabular}{lcccccc}
  \hline\noalign{\smallskip}
  Data set & 1 & 2 & 3 & 4 & 5 & 6 \\
  \hline\noalign{\smallskip}
Area coverage              &0.20\%	&0.99\% &1.55\%   &1.53\%  &1.75\%  &1.99\%\\
Equivalent diameter ($\rm km$)        &181$\pm$22 	&168$\pm$29 	&178$\pm$29 	&195$\pm$36 	&184$\pm$38     &194$\pm$36\\
\quad\quad\quad [\emph{min,max}]      &[\emph{111, 245}]   &[\emph{103, 402}]   &[\emph{109, 440}]   &[\emph{106, 445}]   &[\emph{122, 447}]   &[\emph{112, 473}] \\
Intensity contrast              &0.99$\pm$0.04  &1.01$\pm$0.04  &1.03$\pm$0.04  &1.05$\pm$0.06  &1.05$\pm$0.04  &1.06$\pm$0.05\\
\quad\quad\quad [\emph{min,max}] &[\emph{0.91, 1.12}]   &[\emph{0.90, 1.19}]   &[\emph{0.90, 1.31}]   &[\emph{0.92, 1.30}]   &[\emph{0.92, 1.24}]   &[\emph{0.89, 1.28}]\\
Lifetime ($\rm sec$)            &104$\pm$104     &133$\pm$133     &114$\pm$114     &141$\pm$141     &121$\pm$121     &124$\pm$124\\
\quad\quad\quad [\emph{min,max}] &[\emph{103, 582}]   &[\emph{102, 826}]   &[\emph{120, 723}]   &[\emph{119, 735}]   &[\emph{120, 572}]   &[\emph{114, 580}]\\
Velocity ($\rm km$ $\rm s^{-1}$) &1.35$\pm$0.71  &1.23$\pm$0.64  &1.06$\pm$0.55  &1.04$\pm$0.54  &1.06$\pm$0.55  &1.05$\pm$0.55\\
\quad\quad\quad [\emph{min,max}] &[\emph{0.01, 5.27}]   &[\emph{0, 6.80}]   &[\emph{0.08, 5.43}]   &[\emph{0.02, 5.32}]   &[\emph{0.06, 5.21}]   &[\emph{0.03, 5.75}]\\
Diffusion index                 &1.31$\pm$0.65  &1.21$\pm$0.78  &0.91$\pm$0.43  &1.05$\pm$0.67  &0.86$\pm$0.49  &0.93$\pm$0.77\\
\quad\quad\quad [\emph{min,max}] &[\emph{-3.64, 3.93}]   &[\emph{-4.91, 5.39}]   &[\emph{-4.17, 4.28}]   &[\emph{-5.21, 6.51}]   &[\emph{-7.00, 4.21}]   &[\emph{-5.70, 4.43}]\\
Ratio of motion range           &1.30$\pm$0.80      &1.18$\pm$0.76  &1.11$\pm$0.77  &1.02$\pm$0.62  &0.96$\pm$0.67  &1.03$\pm$0.69\\
\quad\quad\quad [\emph{min,max}] &[\emph{0.31, 5.06}]   &[\emph{0.15, 6.39}]   &[\emph{0.04, 6.28}]   &[\emph{0.17, 5.42}]   &[\emph{0.13, 4.73}]   &[\emph{0.13, 4.79}]\\
Motion type                     &0.69$\pm$0.69  &0.69$\pm$0.69  &0.58$\pm$0.58  &0.59$\pm$0.59  &0.59$\pm$0.59  &0.62$\pm$0.62\\
\quad\quad\quad [\emph{min,max}] &[\emph{0.08, 0.99}]   &[\emph{0.04, 1.00}]   &[\emph{0.23, 0.99}]   &[\emph{0.03, 0.99}]   &[\emph{0, 1.00}]   &[\emph{0.04, 0.98}]\\

  \noalign{\smallskip}\hline
\end{tabular}
\end{center}
\end{table}

  \begin{figure*}
  \centering
   \includegraphics[width=13cm]{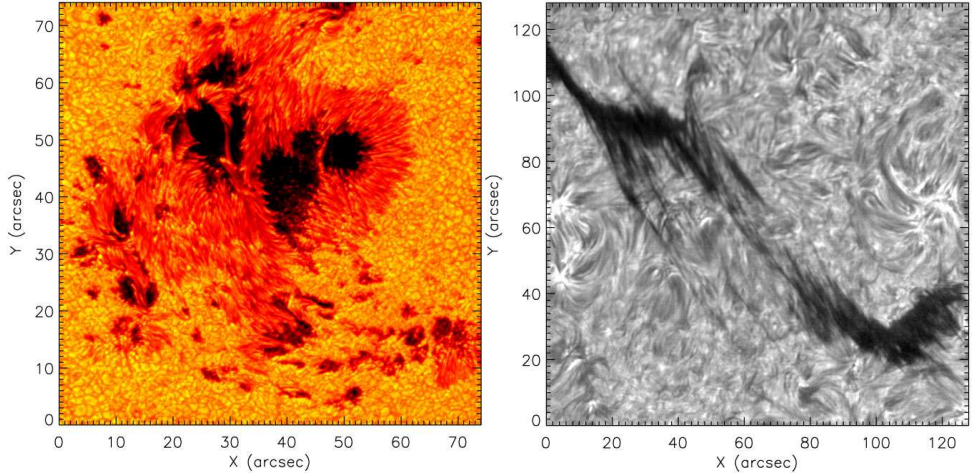}\\
\caption{ {\bf High resolution observation of photosphere and chromosphere.} Left panel: A complex active-region NOAA 12673 in TiO image observed by the NVST at 03:45:18 UT on 2017 September 5. This active-region includes several sunspots and pores. The large sunspots are hosting light bridge. Right panel: Fine structure of a quiescent filament observed at H$\alpha$ line center by the NVST at 05:31:01 UT on 2018 July 23.}
       \label{fig2}
   \end{figure*}
   
  \begin{figure*}
  \centering
   \includegraphics[width=13cm]{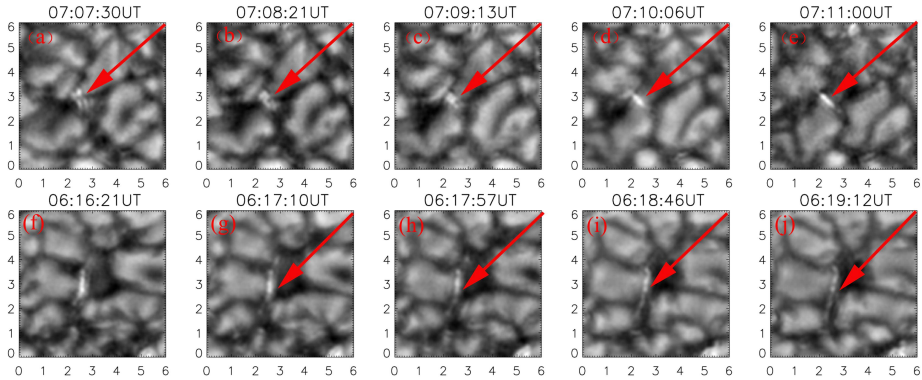}
\caption{{\bf Two different bright points observed by the NVST.} (a-e): Evolution of the isolated bright points (BPs). (f-j):  Evolution of the non-isolated BPs.  Image reproduced with permission from Liu et al. (2018), copyright by AAS.}
       \label{fig2}
   \end{figure*}
   
   \begin{figure*}
  \centering
   \includegraphics[width=7cm]{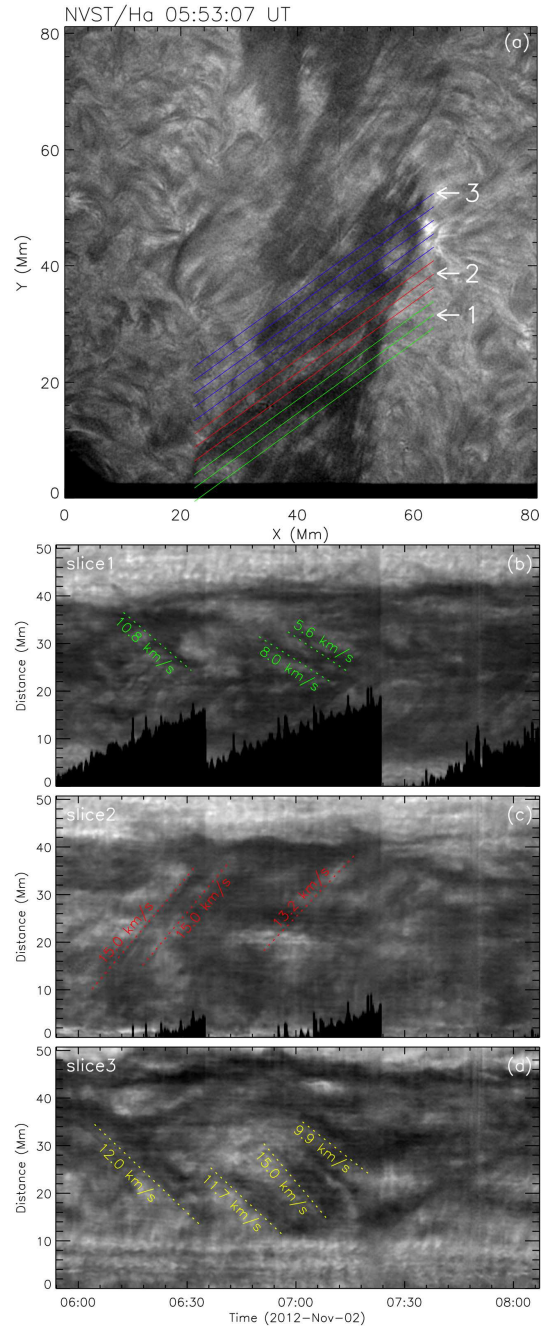}
\caption{{\bf High-resolution H$\alpha$ image of the quiescent filament observed by NVST at 05:53:07 UT on 2012 November 2.} The green and yellow lines mark the areas that have material flows with the same direction and the red lines denote material flows in opposite directions. The lines marked by the numbers indicate the positions of the time slices shown in Fig. 5(b)-5(d). Image reproduced from Yan et al. (2015), copyright by RAA.}
       \label{fig2}
         \end{figure*}
   
          \begin{figure*}
  \centering
  \includegraphics[width=9cm]{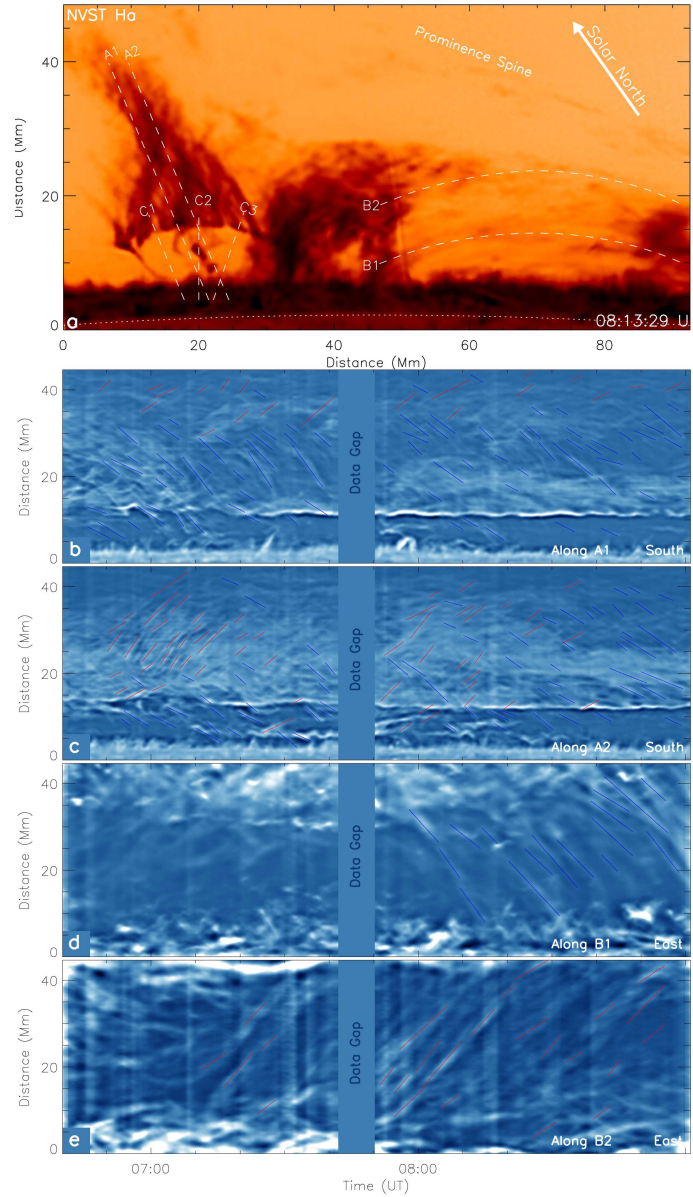}
\caption{{\bf A limb prominence and the counter-streaming} a: A quiescent prominence observed by the NVST. b-e: The time-distance diagrams showing the mass dynamical motion along the slit A1, A2, B1, and B2. Image reproduced with permission from Shen et al. (2015), copyright by AAS.}
       \label{fig2}
   \end{figure*}
   
   \begin{figure*}
  \centering
  \includegraphics[width=13cm]{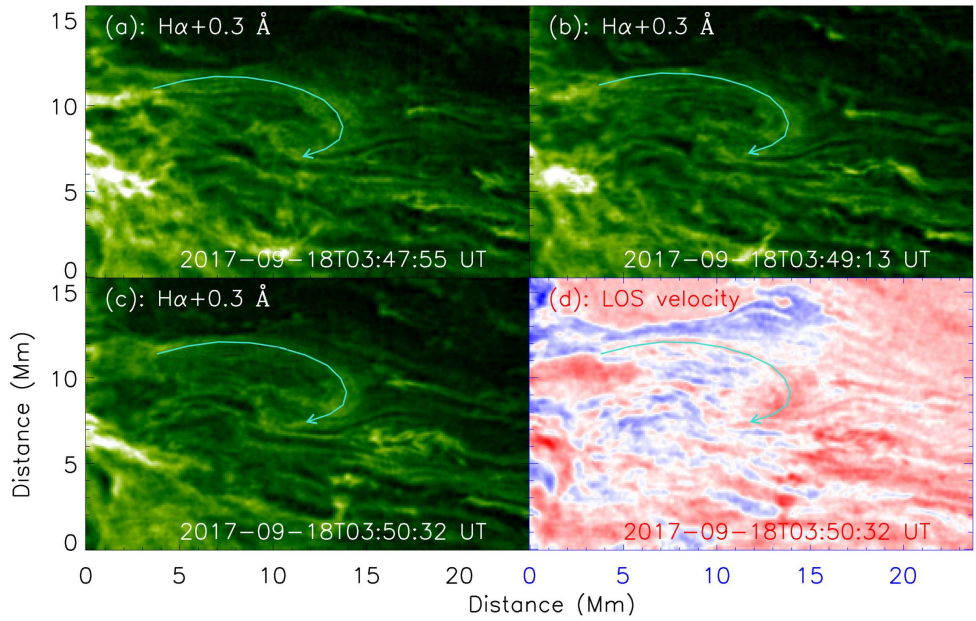}
\caption{{\bf Evolution of a quiescent prominence observed by the NVST in off-band H$\alpha$ image.} a-c: $H\alpha$ +0.3 \AA\ images. d: Line-of-sight velocity. The curved arrows indicate the vortex flow in the quiescent prominence. Image reproduced with permission from Li et al. (2018), copyright by AAS.}
       \label{fig2}
   \end{figure*}

   \begin{figure*}
  \centering
   \includegraphics[width=13cm]{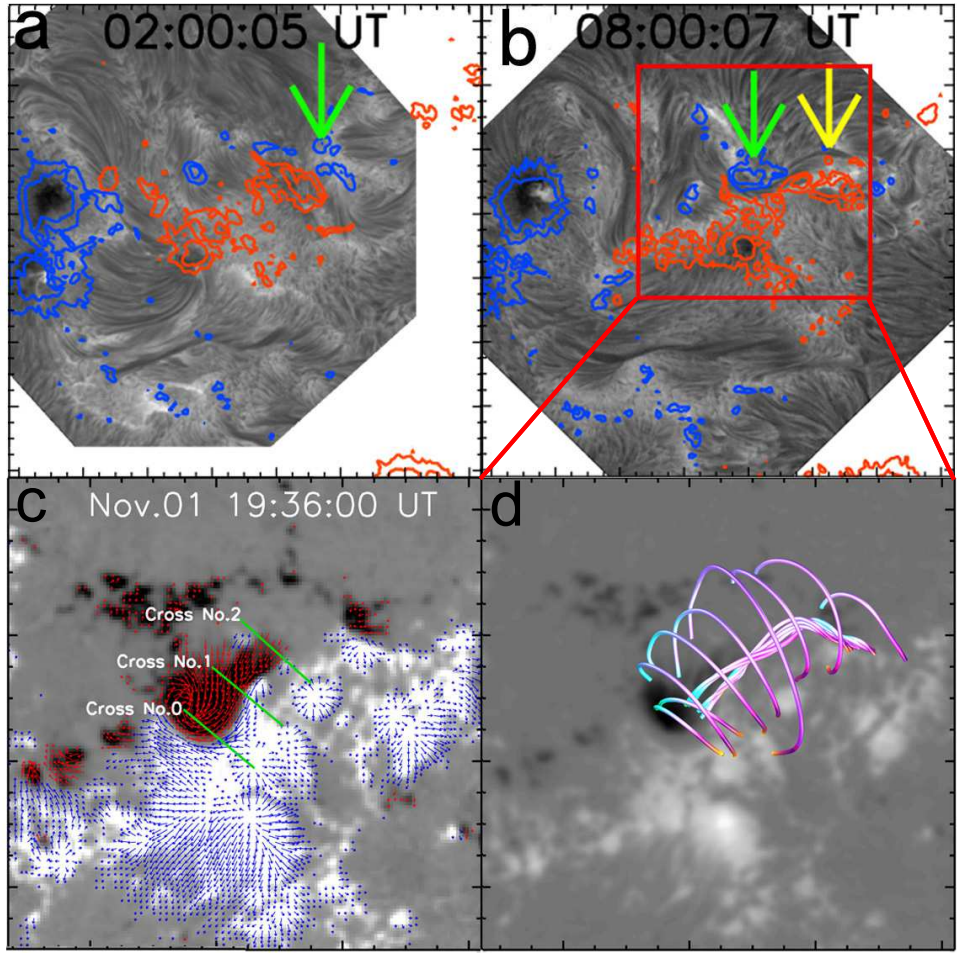}
\caption{{\bf The formation of the active-region filaments observed by the NVST and its magnetic structure.} (a) and (b): $H\alpha$ images observed at 02:00:05 UT on 2013 October 31and at 08:00:07 UT on November 1 by NVST superimposed on the corresponding line-of-sight magnetogram observed by the SDO/HMI. The red and blue contours indicate the positive and negative polarities. The contour levels are $\pm$500 G and $\pm$1000 G. The green and the yellow arrow indicate the small sunspot and the active-region filament. (c): The vector magnetogram observed by SDO/HMI at 19:36:00 UT on November 1. (d): The extrapolation of the filament structure and the surrounding magnetic fields of the filament superimposed on the longitudinal magnetic fields. Image reproduced from Yan et al. (2015), copyright by AAS}
       \label{fig3}
   \end{figure*}

   \begin{figure*}
  \centering
   \includegraphics[width=11cm]{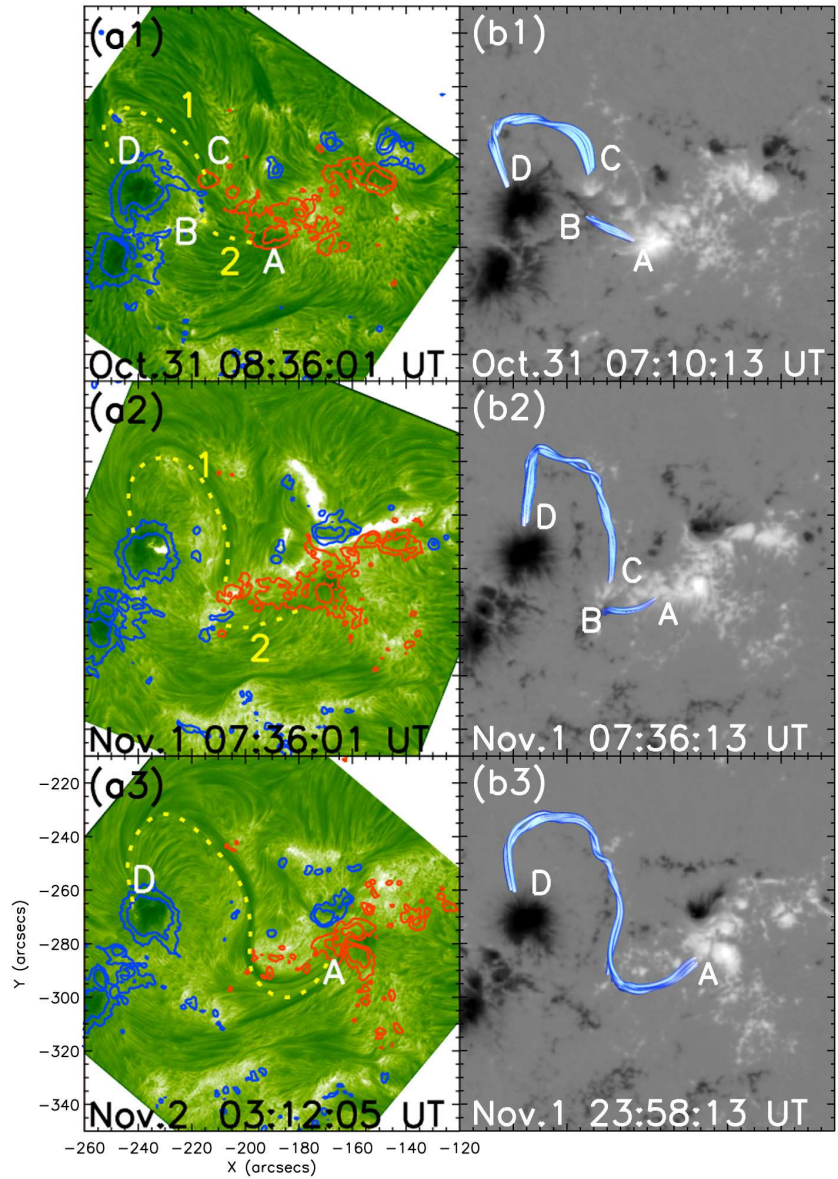}
\caption{{\bf Formation of an active-region filament observed by the NVST and the extrapolation magnetic fields by using NLFFF model.} (a1-a3):  H$\alpha$ images observed by the NVST. (b1-b3): the extrapolations based on the vector magnetograms observed by SDO/HMI. Image reproduced from Yan et al. (2016), copyright by AAS}
       \label{fig2}
   \end{figure*}

   \begin{figure}
  \centering
   \includegraphics[width=14cm]{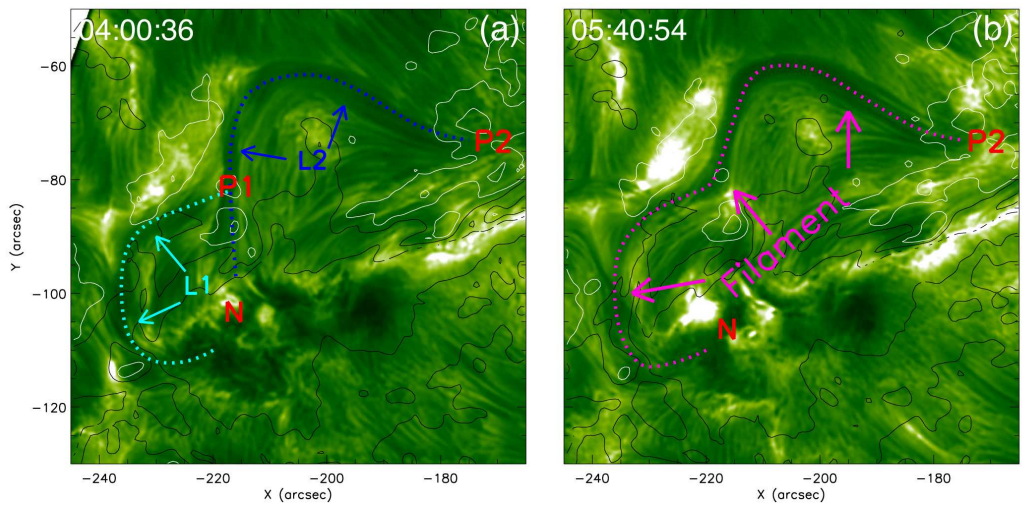}
\caption{{\bf Formation of an active-region filament observed by the NVST.}  L1 and L2 indicate the left and the right part of the chromospheric fibrils marked by the cyan and the blue curved lines in Fig. 8(a), respectively. The pink curved line indicates the newly formed filament. Image reproduced with permission from Xue et al. (2017), copyright by AAS.}
       \label{fig4}
   \end{figure}

\begin{figure*}
  \centering
   \includegraphics[width=13cm]{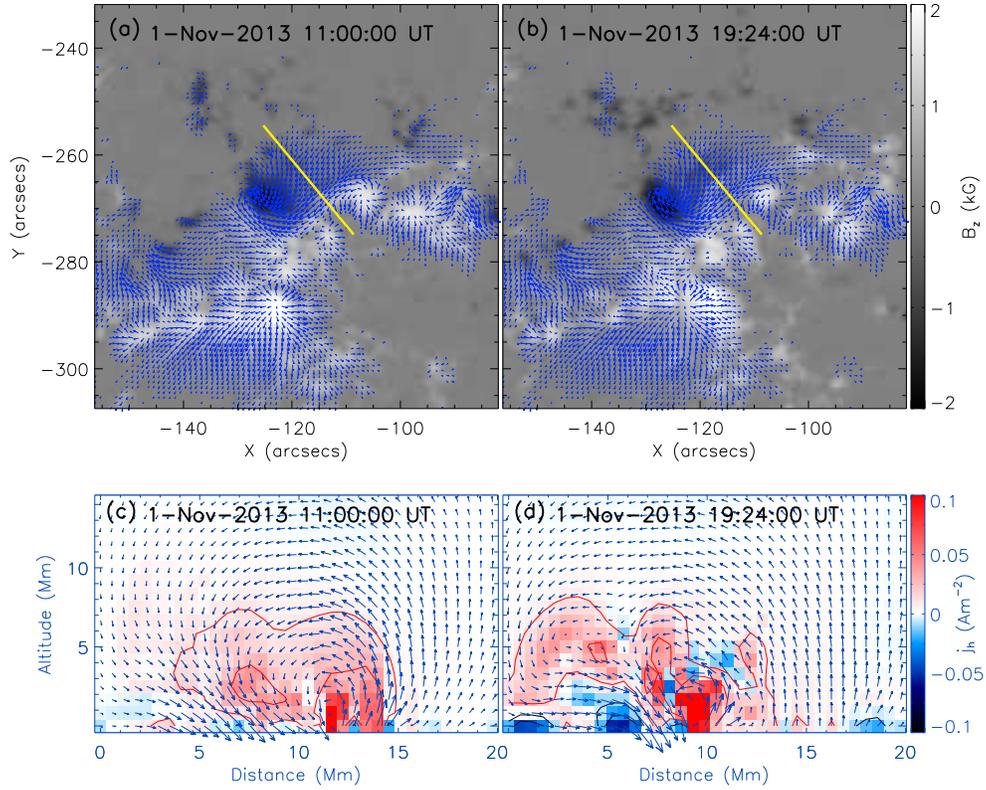}
\caption{{\bf Vector magnetograms and electric currents along the axis of the filaments.} (a-b) Vector magnetograms observed by the SDO/HMI. (c-d) The electric currents along the axis of the filaments derived from the extrapolated 3-D magnetic fields by using NLFFF model. The yellow lines indicate the positions of the cross-section of the electric currents perpendicular to the filaments. Image reproduced with permission from Wang et al. (2016), copyright by AAS.}
     \label{fig1}
   \end{figure*}

    \begin{figure*}
  \centering
   \includegraphics[width=13cm]{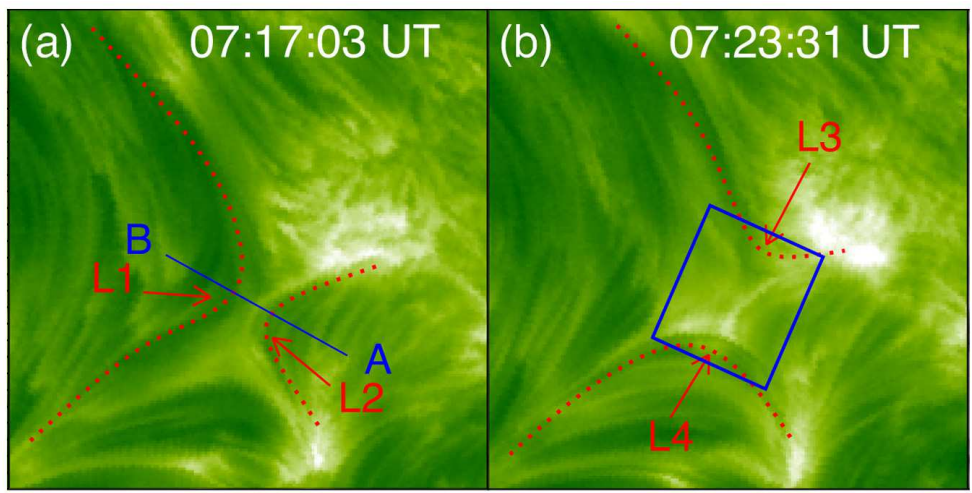}
\caption{{\bf A small-scale magnetic reconnection observed by the NVST.}  L1 and L2 indicate the two groups of the chromospheric fibrils before the magnetic reconnection. L3 and L4 indicate the newly formed chromospheric fibrils after the magnetic reconnection. The blue box indicates the reconnection region. Image reproduced with permission from Yang et al. (2014), copyright by AAS.}
       \label{fig2}
   \end{figure*}

    \begin{figure}
  \centering
   \includegraphics[width=13cm]{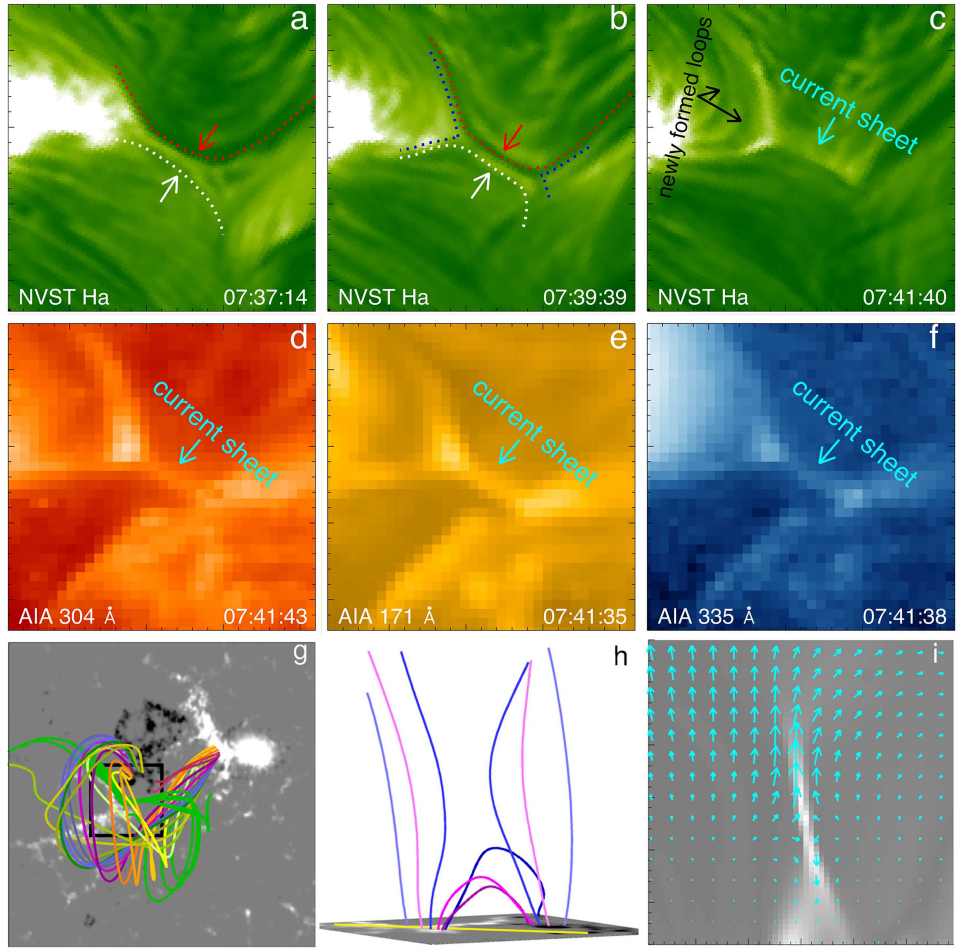}
\caption{{\bf The reconnection process of an active-region filament and the chromospheric fibrils observed by the NVST.} The filament threads and chromospheric fibrils are indicated by red and white arrows respectively in panels (a) and (b). The current sheet forming in magnetic reconnection is marked by the cyan arrows in panels (c)-(f). The results of a data-constrained magnetohydrodynamic simulation reproducing the magnetic reconnection during the filament eruption are presented in panels (g)-(i). Image reproduced with permission from Xue et al. (2016b), copyright by Nature.}
       \label{fig4}
   \end{figure}

    \begin{figure*}
 \centering
   \includegraphics[width=13cm]{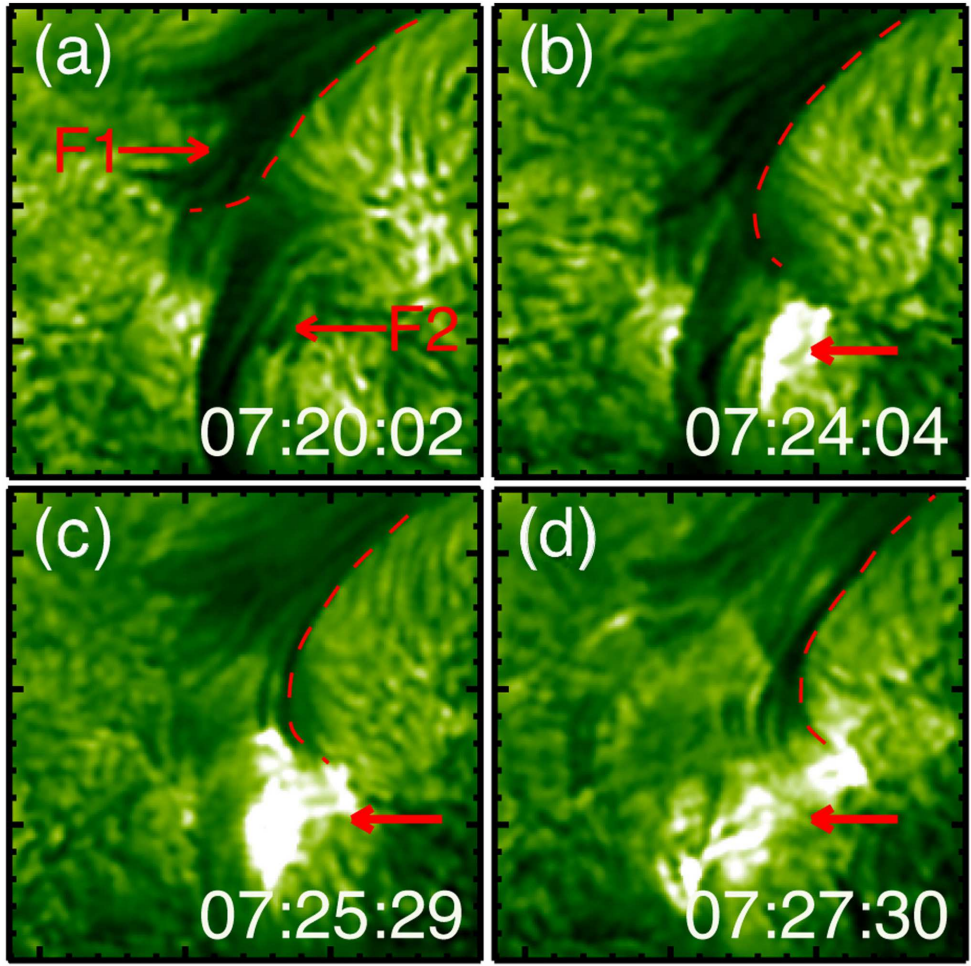}
\caption{{\bf Interaction between two active-region filaments observed by the NVST.}  F1 and F2 indicate the two active-region filaments. The dashed lines in Fig. 12(b)-12(c) outline the newly formed filament threads and the red arrows indicate the brightening caused by the interaction of the two filaments. Image reproduced with permission from Yang et al. (2017), copyright by AAS.}
       \label{fig2}
   \end{figure*}

\end{document}